\documentstyle[12pt,epsf]{article}
\newcommand{\bea}{\begin{eqnarray}}
\newcommand{\eea}{\end{eqnarray}}
\newcommand{\beq}{\begin{equation}}
\newcommand{\eeq}{\end{equation}}
\begin{document}
%\tighten
%\draft
\title{
\Large\bf Coherent States in Field Theory}
\author{
{\bf Wei-Min Zhang}\thanks{e-mail: wzhang@mail.ncku.edu.tw} 
\\
{\small Department of Physics, National Cheng-Kung University,
Tainan, Taiwan 701, R.O.C.}}
%\date{\today}
\maketitle
%\begin{abstract}
\abstract{~~~~Coherent states have three main properties: coherence, 
overcompleteness and intrinsic geometrization. These unique 
properties play fundamental roles in field theory, especially, 
in the description of classical domains and quantum fluctuations 
of physical fields, in the calculations of physical processes 
involving infinite number of virtual particles, in the 
derivation of functional integrals and various effective 
field theories, also in the determination of long-range orders 
and  collective excitations, and finally in the exploration 
of origins of topologically nontrivial gauge fields 
and associated gauge degrees of freedom.}
%\end{abstract}

\normalsize

\section{Introduction}
In the past thirty-six years, the developments and applications 
of coherent states have been made tremendous progress.  Yet, 
the idea of creating a coherent state for a quantum system 
was conceived well before that.  In fact, back in 1926, 
Schr\"{o}dinger first proposed the idea of what is now called 
``coherent states" \cite{Sch26} in connection with the quantum 
states of classical motion for a harmonic oscillator. 
In other words, the coherent states were invented immediately 
after the birth of quantum mechanics.  However, between 1926 
and 1962, activities in this field remained almost dormant, 
except for a few works in condensed matter physics 
\cite{Lee53,And58,Val58} and particle physics \cite{Sch51,Kla60} 
in 50's. It was not until some thirty five years after 
Schr\"{o}dinger's pioneering paper that the first modern 
and systematic application to field theory was made by 
Glauber and Sudarshan \cite{Gla63,Sud63} and launched this 
fruitful and important field of study in theoretical as well as 
experimental physics. 

I became interested in the subject of coherent states about 
fifteen years ago. On the occasion of Prof. Sudarshan visiting 
Suzhou of China (1984), I listened for the first time in life
a topic on coherent states presented by Prof. Sudarshan. 
As a second-year graduate student at that time, I was looking 
for some research problem on collective excitations in strongly 
interacted many-body systems (particularly in nuclear 
physics). Prof. Sudarshan's lecture inspired me to think 
whether under constraint(s) of dynamical symmetries  
collective excitations can be described in terms of coherent 
states, as a result of multi-particle correlations (coherence). 
Later on I realized, this is indeed a very active subject 
covering problems from condensed matter physics to nuclear 
and particle physics. Of course, these coherent states have no 
longer the simple but beautiful form Glauber and Sudarshan 
proposed for light beams. Actually, these states are generated 
by complicated collective composite operators of particle-particle 
pairs or particle-hole pairs. Their mathematical structure 
were already developed in early 70's by Perelomov and Gilmore 
\cite{Per72,Gil72} based on the theory of Lie groups. Newdays, 
the concept of coherent states has been extensively investigated.  
Many methods based on coherent states have also been developed 
for various theoretical problems. Nevertheless, the original 
development of coherent states in quantum electromagnetic 
field (or more precisely, in the study of quantum optical 
coherence) has made tremendous influence in physics. 

One can find that a large body of the literature on coherent 
states has appeared. This vast literature was exhaustively 
collected, catalogued and classified by Klauder and Skagerstam 
\cite{Kla85}. About the mathematical usefulness of coherent
states as a new tool to study the unitary representations of 
Lie groups has been described in a well expository book by 
Perelomov \cite{Per86}.  A review article on the theory of coherent
states and its applications that cover subjects of quantum 
mechanics, statistical mechanics, nonlinear dynamics and many-body 
physics has also been presented by author and his collaborators 
\cite{Zha90}.  In this article, I will only concentrate on the 
topic of coherent states in field theory. As usual, it is 
not my intention to give a complete review about coherent states 
in field theory. An extensive review on coherent states in field 
theory and particle physics may be found in \cite{Ska93}. 
I will rather like to present here a discussion on whether one 
can formulate field theory in terms of coherent states such that 
the new formulation may bring some new insights to the next 
development of field theory in the new millennium.  Coherent 
state can become a useful and important subject in physics 
because of its three unique properties: the coherence, the 
overcompleteness and the intrinsic geometrization. These unique 
properties, in certain contents, are fundamental to field theory. 
I will select some typical topics in field theory that can 
be efficiently described by coherent states based on these 
properties. These topics include the productions of coherent 
states in field theory, the basic formulation of quantum 
field theory in terms of coherent state functional integrals, 
the spontaneously symmetry breaking described from coherent states, 
and the effective field theories derived from coherent states. 
Also, I will ``sprinkle" discussions about the geometrical 
phases of coherent states and their interpretation as gauge 
degrees of freedom in field theory, a subject which has still 
received increasing importance in one's attempt to understand 
the fundamental of nature.	

\section{Photon coherent states}

I may begin with the simplest coherent state of photons, or more 
generally speaking, bosons. Such a set of coherent states has 
been described in most of quantum mechanics text books and are 
familiared to most of physicists.  It is indeed the most popular 
coherent state that has been used widely in various fields.  
The coherent state of photons can describe not only the 
coherence of electromagnetic field, but also many other 
properties of bosonic fields. It is the basis of modern 
quantum optics \cite{Kla68}, and it also provides a fundamental 
framework to quantum field theory, as one will see later.  

%\subsection{Optical Coherence}
By means of {\bf optical coherence}, one may consider the $n$-th 
order correlation function of electromagnetic field:
\beq 
	G^n(x_1,\cdots,x_n,x_{n+1},\cdots,x_{2n}) = {\rm tr}\big\{
		\rho E^-(x_1)\cdots E^-(x_n)E^+(x_{n+1})\cdots
		E^+(x_{2n}) \big\},
\eeq
where $x_i$ is the time-space coordinates, $\rho$ denotes the 
density operator, and $E^\pm(x_i)$ represent the 
electric field operators 
with positive and negative frequency. For simplification, the
polarization of electric field is fixed. According to Glauber
\cite{Gla63} the complete coherence of a radiation field 
is that all of the correlation functions satisfy the following 
factorization condition:
\beq
	G^n(x_1,\cdots,x_n,x_{n+1},\cdots,x_{2n}) = {\cal E}^*(x_1)
	 \cdots {\cal E}^*(x_n){\cal E}(x_{n+1})\cdots{\cal E}(x_{2n}).
\eeq
This condition implies electric field operators must behave 
like classical field variables. It may also indicate the electric 
field operator should have its own eigenstates with the 
corresponding classical field variables as its eigenvalues:
\beq	
	E^+(x_i) |\phi \rangle = {\cal E}(x_i) |\phi\rangle~~,~~
	\langle \phi | E^-(x_i) = \langle \phi |{\cal E}^*(x_i) .
\eeq
Moreover, the density operator must also be expressed in terms 
of the eigenstates $|\phi \rangle$.  Obviously, the 
conventional Fock space in 
quantum theory does not obey the above condition. 

This is actually a nontrivial problem, because it requires  
a complete description of classical motions in terms of quantum 
states. Meantime, the operator $E^\pm(x_i)$ itself is not a 
Hermitian operator. The eigenstate problem of a nonhermitian 
operator is unusual in quantum mechanics.  Fortunately, such 
quantum states have already been constructed by Schr\"{o}dinger 
soon after his invention of quantum mechanics in 1926. In 
order to answer the question how microscopic dynamics transits to 
macroscopic world, Schr\"{o}dinger looked for quantum states which 
follow precisely the corresponding classical trajectories all the 
time, and meantime, the states must also be the exact solution of 
quantum dynamical equation (i.e., the Schr\"{o}dinger equation).  
But only for harmonic oscillator, such states were constructed 
\cite{Sch26}: 
\beq	\label{wpt}
 	\phi_z(x) \sim \exp \Big\{-{1\over 2}(x+z)^2\Big\},
\eeq
where $z$ is a complex variable.
These states are actually the Gaussian wave packets centered on the 
classical trajectory $z=(x+ip)$, $x$ and $p$ are the position of 
harmonic oscillator in the phase space that satisfies classical 
equations of motion. One can show that Eq.~(\ref{wpt}) is 
also an exact solution 
of Schr\"{o}dinger equation. The classicality of Gaussian wave 
packets are manifested by the minimum uncertainty relationship:		
\beq
	\Delta x^2 \Delta p^2 = {\hbar^2 \over 4} ~~~{\rm and}~~~ 
	\Delta p= \Delta x 	.	\label{2unc}
\eeq 
In other words, the wave packets governed by the Hamiltonian of 
harmonic oscillator follow classical trajectories
and do not spread in time . 

Glauder and Sudarshan discovered \cite{Gla63,Sud63} that such 
a wave packet is a superposition of Fock states. It is also an 
eigenstate of $E^+(x_i)$. In quantum field theory, electromagnetic 
field consists of infinite harmonic oscillating modes (photons). 
Explicitly, the Hamiltonian of quantum electromagnetic field 
(in Coulomb gauge) is given by
\beq
H = -{1\over 2} \int d^3 x \Big\{ {\bf E}^2 + {\bf B}^2 \Big\},
\eeq
where ${\bf E}$ and ${\bf B}$ are the electric and magnetic fields. 
The electromagnetic field can be expressed by the vector potential 
${\bf A}: {\bf E} = - \partial{\bf A}/\partial t, {\bf B} = \nabla 
\times {\bf A}$. It is convenient to expand the vector potential 
in terms of plane waves (Fourier series) 
\beq
{\bf A}(x,t) = \int {d^3k\over \sqrt{(2\pi)^3 2\omega_k}} 
	\sum_\lambda  \Big\{ a^\lambda_k \vec{\varepsilon}^\lambda(k) 
	e^{-i k x} + a^{\lambda \dagger}_k \vec{\varepsilon}^{\lambda *}
		(k)e^{i k x} \Big\},
\eeq
where $\vec{\varepsilon}^\lambda(k)$ is the polarization vector of 
electromagnetic field, and $(a^{\lambda\dagger}_k, a^\lambda_k)$ 
are the creation and annihilation operators, 
\beq
	[a^\lambda_k~, ~ a^{\lambda' \dagger}_{k'} ] = 
		\delta_{\lambda \lambda'}\delta_{kk'}, ~~~
	[a^\lambda_k~, ~ a^{\lambda'}_{k'} ] = [a_k^{\lambda\dagger}~, 
		~ a_{k'}^{\lambda' \dagger} ] = 0 .
\eeq
Then the Hamiltonian of electromagnetic field can be deduced to
\beq
	H = \sum_{k\lambda} \omega_k (a^{\lambda\dagger}_k 
		a^\lambda_k + 1/2) ,
\eeq
which means that the electromagnetic field consists of infinite 
individual electromagnetic modes, i.e., photons. Each 
photon corresponds to a harmonic oscillator.
 
In the particle number representation, the Gaussian wave packet  
can be written as
\beq
	|z\rangle = \exp(-{1\over 2}|z|^2) \exp(za^\dagger)|0\rangle .
\eeq
where $|0\rangle$ is the vacuum state: $a |0\rangle=0$. From the
above expression, it is easy to show that the wave packet is also 
an eigenstate of the annihilation operator $a$: 
\beq
	a | z \rangle = z | z \rangle ,	
\eeq
Thus, the quantum state describing the optical coherence of 
electromagnetic field can be expressed by
\beq	\label{ptcs}
	|\{z^\lambda_k\} \rangle = \exp\Big\{-{1\over 2}
		\int d^3 k \sum_\lambda |z^\lambda_k|^2\Big\} 
		\exp\Big\{\int d^3 k \sum_\lambda 
		z^\lambda_k a^{\lambda\dagger}_k \Big\}|0\rangle ,
\eeq
which is an eigenstate of the positive frequency part of the 
electric field operator,
\beq
	{\bf E}^+(x)|\{z^\lambda_k\} \rangle = {\bf\cal E}(x)
		|\{z^\lambda_k\} \rangle ~~,~~~
		{\bf\cal E}(x)=i \int {d^3 k\over (2\pi)^{3/2}} 
		\sqrt{\omega_k\over2} \sum_\lambda z^\lambda_k 
		\vec{\varepsilon}_\lambda(k) e^{-i (\omega_k t -
		{\bf k}\cdot {\bf x})} .
\eeq
Besides, the above state has another very important property:
 it supports the following resolution of identity:
\beq	\label{cscs}
	\int |\{z^\lambda_k\}\rangle \langle \{z^\lambda_k\} | 
		\prod_{k\lambda}{dz^\lambda_kdz^{\lambda*}_k 
		\over 2\pi}= I.
\eeq
In other words, these states in the complex space (in terms 
of the variable $z$) form a complete set of states 
(more precisely speaking, 
it is overcomplete because of the continuity of these states). 
This complete set is certainly very different from the set of 
Fock states. Because of the overcompleteness and the analyticity 
of these states, one can expand the density operator by 
(\ref{ptcs}) in a diagonal form
(the so-called P-representation \cite{Gla63}):
\bea
	\rho &=& \int P(\{z^\lambda_k\})|\{z^\lambda_k\}\rangle 
		\langle \{z^\lambda_k\} | \prod_{k\lambda}
		dz^\lambda_k dz^{\lambda*}_k ,\nonumber \\ 
	& & {\rm tr}\rho = \int P(\{z^\lambda_k\})\prod_{k\lambda}
		dz^\lambda_k dz^{\lambda*}_k=1 .
\eea
where $P(z)$ is a weight function. In terms of these states 
(\ref{ptcs}), the factorization criterion of coherent light 
beams is automatically satisfied. Glauber named such states the 
{\it coherent states}. To be more specific, one may call them the 
``photon coherent states".  Physically, the photon coherent states 
have a well-defined phase for each mode. Therefore, coherent  
light beams can be completely described in quantum mechanics in 
terms of photon coherent states.  For these who wish to have 
more detailed discussion on physical consequences of the photon 
coherent states in quantum optics, please refer to the excellent
book by Klauder and Sudarshan \cite{Kla68}.

\section{Coherent states and $S$-Matrix}
As we have seen, the photon coherent state was introduced by 
the requirement of optical coherence. 
Here, I may ask a more general question, namely, how are photon
coherent states generated in field theory? In field theory,
all physical quantities are derivable from the vacuum-to-vacuum
transition amplitude in the presence of external sources. 
It can show that the final state in such processes is a 
coherent state if there is no other interactions except 
for a linear interaction with the
external field. 

To be specific, one may consider the electromagnetic field 
interacting with a classical source:
\beq
	{\cal L} = - {1\over 4}F_{\mu \nu}F^{\mu \nu} - 
		 A_\mu j^\mu,
\eeq
where the classical source is a conserved current: 
$\partial^\mu j_\mu=0$. In the Feynman gauge, the 
equation of motion for the electromagnetic
field is given by
\beq
	\partial_\mu F^{\mu \nu} = \Box A^\nu = j^\nu.
\eeq
A general solution of the above equation is
\beq
	A^\mu(x) = A^\mu_0(x) + \int d^4 y \Delta(x-y) j^\mu (y),
\eeq
where $A^\mu_0(x)$ is the solution of free field, and 
$\Delta(x-y)$ is the Green function determined by 
$\Box_x \Delta(x-y)=\delta^4 (x-y)$. If one assumes 
that the interaction is switched on adiabatically in
a finite time interval, then
\bea	
A^\mu(x) &=& A^\mu_{\rm in}(x) + \int d^4 y \Delta_{\rm ret}
		(x-y) j^\mu (y) \nonumber \\
	&=& A^\mu_{\rm out}(x) + \int d^4 y \Delta_{\rm adv}(x-y) 
		j^\mu (y), \label{asps}
\eea
where the retarded and advanced Green functions are given by
\beq
	\Delta_{^{\rm ret}_{\rm adv}}(x) = -{1\over (2\pi)^4} \int 
		d^4p {e^{-ipx}\over (p_0\pm i\epsilon)^2 -{\bf p}^2},
\eeq
and $A^\mu_{\rm in}$ and $A^\mu_{\rm out}$ are free fields 
describing the photon field before and after its interaction 
with the classical course $j^\mu$. The corresponding photon 
states are the in- and out-states (denoted by 
$|\rangle_{\rm in}$ and $|\rangle_{\rm out}$, respectively). 
The in- and out-states form two complete sets of 
free states constructed as a Fock space by the free field operators
$A^\mu_{\rm in}$ and $A^\mu_{\rm out}$.  Therefore, there must 
exist a unitary transformation $S$ (namely $S$-matrix) to 
connect these two complete sets: 
\beq
	A^\mu_{\rm out} = S^\dagger A^\mu_{\rm in}S~~,~~~
		|\rangle_{\rm out} = S^\dagger |\rangle_{\rm in}.
\eeq
From (\ref{asps}), one can see that
\bea
	A^\mu_{\rm out}(x) &=& A^\mu_{\rm in}(x) + \int d^4 y 
		[\Delta_{\rm ret}(x-y)-\Delta_{\rm adv}(x-y)] 
		j_\mu (y) \nonumber \\
	&=& A^\mu_{\rm in}(x)+A^\mu_{\rm cl}(x),
\eea
and $A^\mu_{\rm cl}(x)$ is a $c$-number (classical) field 
generated by the classical current $j^\mu(x)$. Notice that 
$\Delta_{\rm adv}(x)- \Delta_{\rm ret}(x)=\Delta(x)$ which 
relates to the commutator of free fields $[A_{\rm in}^\mu(x),
A_{\rm in}^\nu(y)]=[A_{\rm out}^\mu(x),
A_{\rm out}^\nu(y)]=-ig^{\mu \nu}\Delta(x-y)$.
One may check that the $S$-matrix can be written as
\beq
	S = \exp \Big\{-i \int d^4x A_{\rm in} \cdot j(x)\Big\}
		= \exp \Big\{-i \int d^4x A_{\rm out} \cdot j(x)\Big\}.
\eeq
If we start at time $-\infty$ from the vacuum state 
$|0\rangle_{\rm in}$,
the final state after the free field $A^\mu(x)$ interacted
with the classical current $j^\mu(x)$ becomes a coherent state:
\beq
	|0\rangle_{\rm out} = \exp \Big\{i\int d^4x A_{\rm in}
		(x)\cdot j(x)\Big\} |0\rangle_{\rm in} .
\eeq 
In terms of the Fourier expansion, 
\beq
	A_\mu(x) = \int {d^3k \over \sqrt{2(2\pi)^2 |{\bf k}|}} 
	\sum_\lambda \Big\{ a^\lambda_k \varepsilon_\mu^\lambda(k) 
		e^{-i k x} + a^{\lambda\dagger}_k 
		\varepsilon^{\lambda*}_\mu(k) e^{i k x} \Big\}.
\eeq
The final state can be expressed as 
\beq
|0\rangle_{\rm out} = \exp \Big\{-{1\over 2} \int d^3 k \sum_\lambda 
	|z^\lambda_k|^2 \Big\} \exp \Big\{ \int d^3k \sum_\lambda 
	z^\lambda_k a^{\lambda \dagger}_k \Big\} |0\rangle_{\rm in} 
		=|\{z_k^\lambda\}\rangle ,	\label{csgcs}
\eeq
where $z^\lambda_k= \varepsilon^\lambda(k)\cdot j(k)$, and $j^\mu(k)$ 
is the Fourier transform of the classical current $j^\mu(x)$. This is 
the same photon coherent states introduced by Glauber in the study of  
quantum optical coherence. 

Indeed, one can derive similarly the photon coherent state for the 
laser beams (discussed in the previous section) from a more 
microscopic picture. The Hamiltonian in quantum optics that 
describes the interaction between $N$ atoms and the 
electromagnetic field can be written as:  		
\beq
	H = \sum_k \omega_k a^\dagger_k a_k + \sum_i
		\epsilon_i \sigma_0^i + \sum_{k,i} \gamma_{ki}(t)
		\Bigg\{ {\sigma_+^i \over \sqrt{N} } a_k +  {\sigma_-^i 
		\over \sqrt{N} } a_k^\dagger \Bigg\},  \label{2h1}
\eeq
where $\gamma_{ki}$ are the coupling coefficients between atoms 
and electromagnetic field.  One of the crucial assumptions 
made in the construction of the above Hamiltonian for laser beams 
is that each of the $N$ atoms, labeled by the index $i$, is a 
two-level system and therefore its dynamical variables are  
the usual Pauli operators $\{\sigma_0^i, \sigma_+^i, \sigma_-^i \}$. 
Furthermore, the atomic variables can be treated as a classical 
source (i.e. the spin operators $\sigma^i$ can be regarded as 
$c$-numbers), and the coupling strength $\gamma_{ki}$ are identical
for all the atoms (i.e. $\gamma_{ki}=\gamma_k$). Then,
Eq.~(\ref{2h1}) is reduced to
\bea
	&& H = \sum_k  \omega_k a^\dagger_k a_k + \sum_i
		\epsilon \langle \sigma_0^i \rangle
		+ \sum_{k,i} \gamma_{ki}(t) \Bigg\{ {\langle \sigma_+^i 
		\rangle \over \sqrt{N} } a_k +  {\langle \sigma_-^i 
		\rangle \over \sqrt{N} } a_k^\dagger \Bigg\} \nonumber \\
	&& ~~~ = \sum_k \omega_k a^\dagger_k a_k + \sum_k 
		\Big[\lambda_k(t) a_k^\dagger  + \lambda_k^*(t) a_k \Big]
		 + {\rm const.} 
\eea
This corresponds to the electromagnetic field interacting with an
external time dependent source. The general solution of the 
Schr\"{o}dinger equation for this Hamiltonian is the photon
coherent states. This provides the microscopic picture how 
the photon coherent state is generated and why it becomes the 
fundamental of quantum optics.

Soon after the development of coherent states in optical 
coherence, it was found that the photon coherent state also plays 
an important role in solving the {\bf infrared divergence}  
in quantum electrodynamics for electron scatterings 
\cite{Cha65,Gre67,Kib68,Kul70,Zwa75} (also see the review by
Papanicolaou \cite{Pap76}). As it is well known, the matrix
element in quantum electrodynamics for the scattering of an initial 
state containing a finite number of electrons and photon into a similar
final state has a logarithmical infrared divergence for the small
momentum $k$ \cite{Yen61}. This is because in an actual scattering 
experiment, electromagnetic fields interact with the source
particles so that an infinite number of soft photons are emitted.
These emitted soft photons form a coherent state to the final state, 
as we have discussed. To be more specific, consider a single electron 
scattering.  The source particle can be represented by a classical
current. The Fourier transform of the classical current is given
by 
\beq
	j^\mu(k) = {ie \over \sqrt{2(2\pi)^3 |{\bf k}|}} \Big(
		{p_{f}^{\mu}\over p_f\cdot k} - 	{p_{i}^{\mu}\over 
		p_i\cdot k} \Big),
\eeq
where $p_{i,f}$ are the electron's momentum in the initial and final 
states. If one sums the cross sections over all possible 
final states containing any number of soft photons with momenta 
below the threshold of observability [by using the photon coherent state 
(\ref{csgcs}) with the above classical source], the infrared divergence 
is canceled. This gives a beautiful solution to the infrared 
problem in quantum electrodynamics.

Moreover, one can also show that if the matrix element for scatterings 
are calculated with the initial and final states containing infinite 
number of soft photons by the photon coherent state, the 
infrared divergences are canceled order by order at matrix element
level (not only in cross sections) \cite{Cha65}. The photon coherent 
stat may also use to remove the similar infrared problem in quantum 
gravity, as noticed by Weinberg \cite{Wei65}.  These are perhaps the 
second important applications of the photon coherent state in field 
theory. In addition, one has also attempted to use coherent
states to treat infrared divergences in non-abelian gauge theory
\cite{Eri82}. However, in the non-abelian gauge theory, the infrared 
divergence is much more complicated \cite{Lee64,Ste96}. It contains
two-type infrared divergences, the massless soft infrared divergence
and collinear divergence. It is not clear whether one can construct
some non-abelian coherent states to handle both the soft and collinear
infrared divergences.
 
\section{Functional integrals in field theory}
When the quantum fields interact with quantum fields rather than 
classical external fields, the $S$-matrix (or the time-evolution operator)
does not generate coherent states from the incoming vacuum. In such 
cases, coherent states are useful in the derivation of 
functional integral in field theory. Quantum field 
theory can be reformulated in terms of coherent states not only 
because of its classicality and being eigenstates of the annihilation 
operator.  As we have mentioned, the coherent states are overcomplete:
\beq	\label{cscs2}
	\int |z\rangle \langle z | {dzdz^* \over 2\pi}= I .
\eeq
All these three properties (the classicality, the eigenstates of the
positive frequency part of field operator and the overcompleteness)
together allow one to reformulate quantum field theory in terms of 
a functional integral. Actually, the content of this section can be 
found in many text books, but for completeness I will repeat these 
discussions here.

The ordinary path integral of quantum mechanics developed by Feynman 
\cite{Fey48} can be obtained from the evolution operator by writing 
the evolution operator as 
\beq
	U(t_f, t_0) = \exp\Big\{ -i H(t_f-t_0) \Big\}
		= \lim_{N\rightarrow \infty} \Big(\exp{ -\Big\{i 
		H{t_f-t_0 \over N}\Big\}}\Big)^N	\label{eopd}
\eeq
and then inserting a resolution of the identity in terms of the position
states 
\beq
	\int dx | x \rangle \langle x| =I	\label{csxs}
\eeq
between the terms of above product.  This results in the 
familiar path integral of quantum mechanics,
\bea
	\langle x'(t_f) | x(t_0) \rangle &=& \langle x' 
		|U(t_f, t_0)|x \rangle \nonumber \\
		&=& \int [dx(t)] \exp \Big\{ i 
	\int_{t_0}^{t_f} dt {\cal L}(x(t), \dot{x}(t)) \Big\},
\eea
where ${\cal L}$ is a classical Lagrangian which 
generally has a form of
\beq
	{\cal L}(x, \dot{x}) = {1\over 2} \Big( 
		{dx \over dt}\Big)^2 - V(x),
\eeq
and $[dx(t)] \equiv \prod_{t_0 \leq t \leq t_f} dx(t)$ 
is a functional measure of the path integration \cite{Fey65}. 

Instead of using the basis of the position eigenstates (\ref{csxs}), 
we may use the coherent state basis and insert the resolution of 
identity (\ref{cscs2}) between the terms of product (\ref{eopd}). 
Then a phase space formulation of path integrals can be obtained 
as was first proposed by Klauder \cite{Kla60,Kla79} (More detailed 
derivation will be given later in the application to field theory),
\bea
 	\langle z'(t_f) | z(t_0) \rangle &=& \langle z' |U(t_f, t_0)
		|z \rangle \nonumber \\
		&=& \int [dx(t)][{dp(t)\over 2\pi}] \exp \Big\{ i 
		\int_{t_0}^{t_f} dt {\cal L} (x(t), p(t)) \Big\}, 
		\label{2cspi}
\eea
with
\bea
 	{\cal L} (x, p) &=& \langle z | i {d \over dt} |z\rangle 
		- \langle z |H|z \rangle \nonumber \\
	&=&  {1 \over 2}\Big(p {dx\over dt} - x
		{dp \over dt}\Big)  - {\cal H} (x, p) . 
\eea
where $z=(x+ip)/\sqrt{2}$ and $z^*=(x-ip)/\sqrt{2}$, with the 
initial and final positions $x(t_0)$ and $x(t_f)$ fixed.
This derivation of Feynman's path integral is particularly 
useful in obtaining a functional integral of quantum field 
theory. 

To derive a functional integral of quantum field theory, 
we may start with a neutral scalar field $\phi(x)$, for 
simplification. The Lagrangian density of a neutral 
field is given by
\beq
	{\cal L} = {1\over 2}[(\partial \phi)^2 - m^2\phi^2] 
-V(\phi) ,
\eeq
where $V(\phi)$ represents a self-interacting potential, 
such as ${\lambda\over 4!}\phi^4(x)$. The canonical momentum 
density  conjugate to $\phi({\bf x},t)$ is determined by 
$\pi({\bf x},t)= \partial {\cal L}/\partial \dot{\phi}
({\bf x},t)$. Then the canonical quantization 
leads to
\beq
	[\phi_{op}({\bf x},t), \pi_{op}({\bf x}',t)]= i 
	\delta^3({\bf x-x'}),
\eeq
In the plane-wave expansion, one has
\bea
	 \phi_{op}({\bf x},t)&=&\int {d^3 k \over (2\pi)^{3/2}}
		{\sqrt{1\over 2\omega_k}} \Big\{ a_k  e^{-i k x} 
		+ a^\dagger_k e^{i k x} \Big\}, \nonumber \\
	\pi_{op}({\bf x},t)&=& -i \int {d^3 k \over (2\pi)^{3/2}}
		\sqrt{\omega_k\over 2} \Big\{ a_k  e^{-i k x} 
		- a^\dagger_k e^{i k x} \Big\},
\eea
and the quantum Hamiltonian can be written as  
\beq
	H(t) = \int d^3{\bf x}:\Big\{{1\over 2}[\pi_{op}^2
		+ (\bigtriangledown\phi_{op})^2 +m^3\phi^2_{op}] 
		+V(\phi_{op})\Big\}:
\eeq
here $:~:$ denotes the normal ordering with respect to the 
creation and annihilation operators $a^\dagger_k$ and $a_k$.  
Now, one can define the scalar field coherent state as
\bea
	|\phi\pi\rangle &=&  \exp \Big\{i\int d^3 x [\pi(x)
		\phi_{op}(x) - \phi(x)\pi_{op}(x)]\Big\} |0\rangle 
		\nonumber \\
	&=& \exp \Big\{-{1\over 2}\int d^3 k~ |z_k|^2 \Big\} \exp 
		\Big\{\int d^3 k ~(z_k a^\dagger_k) \Big\} |0\rangle  ,
\eea
from which a functional integral of field theory can be  
derived explicitly. Note that the coherent state in field
theory is defined at a given instant time $t$  
over the whole space $\{{\bf x}\}$ . 

Since the Hamiltonian formalism of field theory is the same as in 
quantum mechanics, one can directly calculate 
the time Green's function 
defined as the matrix element of the evolution operator 
in coherent state basis,
\beq
G(t_f,t_0) = \langle \phi' \pi' | U(t_f, t_0) | \phi \pi \rangle  
	= \langle \phi' \pi' |T\exp\Big\{ -i\int_{t_0}^{t_f} dt H(t) 
		\Big\} | \phi \pi \rangle ,	\label{6pkho}
\eeq
where $T$ is the time-ordering operator. One may slice 
the time interval $t_f-t_0$ into $N$ equal segments: $\varepsilon 
= (t_f-t_0)/N$ so that in the sense of $N \rightarrow \infty$, 
the evolution operator can be written as a subsequently 
multiplication 
of the evolution operator in the interval $\varepsilon$:
\bea
 U(t_f, t_0) &=& \exp \Big\{ -i \varepsilon H(t_n) \Big\}\exp 
		\Big\{ -i \varepsilon H(t_{n-1})\Big\} \nonumber \\
	&& ~~~~~~~~~~~~~~~~
	\cdots \exp \Big\{ -i \varepsilon H(t_2) \Big\}\exp 
		\Big\{ -i \varepsilon H(t_1) \Big\}.
\eea
Using the same procedure as in the derivation of Feynman's path 
integral in quantum mechanics, one should insert the resolution 
of identity,
\beq
	 \int [d\phi({\bf x})][{d\pi({\bf x})\over 2\pi}]|\phi 
		\pi\rangle \langle \phi \pi| = I, ~~~ 
\eeq 
at each interval point, where $[d\phi({\bf x})]\equiv \prod_{-\infty 
<{\bf x}<\infty}d\phi({\bf x})$, etc. are defined over the whole space. 
Then  
\beq
	G(t_f,t_0) = \lim_{N \rightarrow \infty} \int \Bigg(\prod_{i=1
		}^{N-1} [d\phi_i({\bf x})][{d\pi_i({\bf x})\over 2\pi}]
		\Bigg)\prod_{i=1}^N \langle \phi_i \pi_i|\exp\Big\{-i
		 \varepsilon H(t_i)\Big\} |\phi_{i-1}\pi_{i-1} \rangle .
\eeq
Up to the first order in $\varepsilon$, 
\beq
\langle \phi_i \pi_i | \exp \left( - i \varepsilon H(t_i)
\right)| \phi_{i-1} \pi_{i-1})\rangle \approx \langle \phi_i \pi_i|
		\phi_{i-1} \pi_{i-1}\rangle \exp \left(-i \varepsilon 
	{\langle \phi_i \pi_i |H(t_i)|\phi_{i-1} \pi_{i-1}\rangle 
	\over \langle \phi_i \pi_i | \phi_{i-1} \pi_{i-1} \rangle} 
		\right) .
\eeq
Note that the coherent state $|\phi \pi \rangle$ is normalized. 
In the limit of $\varepsilon \rightarrow 0 $ (i.e. $N \rightarrow 
\infty$), 
\bea
	\langle \phi_i \pi_i| \phi_{i-1} \pi_{i-1}\rangle &=& 1 - 
	\langle \phi_i \pi_i| (|\phi_i \pi_i \rangle -|\phi_{i-1} 
		\pi_{i-1}\rangle) \nonumber \\
	&\simeq&  \exp \Big\{ i \varepsilon \langle \phi_i \pi_i
		| i {\Delta |\phi_i \pi_i \rangle \rangle 
		\over \varepsilon} \Big\} ,
\eea
where $\Delta |\phi_i \pi_i \rangle \equiv |\phi_i \pi_i \rangle -
|\phi_{i-1} \pi_{i-1}\rangle$. Then, the Green's function becomes
\bea
 	G(t_f,t_0) &=& \lim_{N \rightarrow \infty} \int 
		\Bigg(\prod_{i=1}^{N-1} [d\phi_i({\bf x})][{d\pi_i({\bf x})
		\over 2\pi}]\Bigg) \nonumber \\
	& & ~~~~~~~~~~~~ \times \exp i \sum_{i=1}^N \varepsilon \Big\{ 
		\langle \phi_i \pi_i| i {\Delta |\phi_i \pi_i \rangle 
		 \over \varepsilon}  - \langle \phi_i 
		\pi_i | H(t_i) |\phi_i \pi_i \rangle \Big\} \nonumber \\
&=& \int [d\phi(x)][{d\pi(x)\over 2\pi}] \exp \Big\{ i \int_{t_0}^{t_f} 
		dt \Big[\langle \phi \pi | i{d\over dt} |\phi \pi 
		\rangle- \langle \phi \pi | H |\phi \pi 
		\rangle \Big] \Big\} \nonumber \\
&=& \int [d\phi(x)][{d\pi(x)\over 2\pi}] \exp \Big\{ i \int_{t_0}^{t_f} 
		dt \int d^3 x \Big[{1\over 2}(\pi \dot{\phi}-\phi 
		\dot{\pi}) - {\cal H}(x)\Big] \Big\} , \label{2cspii}
\eea
with
\beq
 	{\cal H} = {1\over 2}[\pi^2 + (\bigtriangledown\phi)^2 
		+m^3\phi^2] +V(\phi) .
\eeq
As we see that the coherent state gives a natural derivation
of path integrals in field theory.  

In field theory, the correlations between $n$ fields are defined
by the $n$-point Green functions,  
\beq
	G^{(n)}(x_1,\cdots,x_n) = \langle 0|T( \phi(x_1) 
		\cdots \phi(x_n))|0\rangle ,
\eeq
which can be determined from the generating functional $W(J)$ which
 is defined as the vacuum-to-vacuum amplitude in the presence 
of external current $J(x)$:
\beq
	W(J) =\langle 0|U(-\infty, \infty)|0\rangle_J .
\eeq
This generating functional can then be expressed in terms of the 
time Green's function $G(t_f,t_0)$ by adding a term $\int d^3 x 
J(x)\phi(x)$ 
in the exponent and then taking $t_0 \rightarrow -\infty$ and $t_f 
\rightarrow \infty$:
\bea
 W(J) &=& \int [d\phi(x)][{d\pi(x)\over 2\pi}] \exp \Big\{\int d^4 x 
	\Big[{1\over 2}(\pi \dot{\phi}-\phi \dot{\pi}) - {\cal H}(x) 
		+ J(x)\phi(x)\Big] \Big\} \nonumber\\
	 &=& \int [d\phi(x)] \exp \Big\{\int d^4 x \Big[{\cal L}(x) 
		+ J(x)\phi(x)\Big] \Big\} = \exp{iZ(J)}, 
\eea
and $Z(J)$ is a functional partition function in quantum field 
theory. The $n$-point Green's functions containing only the connected 
graphs is given by,
\beq
G_c^{(n)}(x_1,\cdots,x_n)= {(-i)^n\over W(J)} {\delta^n W(J) \over
	\delta J(x_1) \cdots \delta J(x_n)}\Bigg|_{J=0}= (-i)^{n-1}
		{\delta^n Z(J) \over \delta J(x_1) \cdots \delta J(x_n)}
		\Bigg|_{J=0}.	
\eeq

Taking the stationary phase approximation of $W(J)$ naturally 
results in the classical equations of motions \cite{Fed76,Kla79}. 
On the other hand, after integrating out the 
$\pi(x)$ field, the functional integrals $W(I)$ and $Z(J)$ become
covariant. Now, all physical quantities in field theory are, 
in principle, derivable from $W(J)$ or $Z(J)$ in a covariant form, 
which are standard in text books. I should not repeat these 
discussion here. 

The above formulation is only for bosonic fields. Field theory that 
describes the real world must also involve fermion (matter) 
fields. To formulate a similar functional integral for 
fermionic fields, one needs to introduce fermion coherent 
states. Similarly, one may try to construct
such a coherent state as an eigenstate of the fermion annihilation
operator:
\beq
	c_i |\xi \rangle = \xi_i | \xi\rangle~~, 
		~~~\{c_i, c^\dagger_j \}=\delta_{ij} . 
\eeq
However, since the fermion creation and annihilation operators 
satisfy the anticommutation relationship, the eigenvalue 
$\xi_i$ of the annihilation operator is its classical 
analogy which cannot be an ordinary number. The quantum-classical 
corresponding principle simply requires that $\xi_i$
must be anticommute:
\beq
\xi_i \xi_j = - \xi_j \xi_i~~, ~~~\xi_i \xi^*_j = \xi^*_j \xi_i~~,
		~~~\xi^2_i = 0 .
\eeq
The numbers satisfies the above relations are called Grassmann 
numbers. Functions of Grassmann numbers are given by
\beq
f(\xi_i) = f^0+f^{(1)}_i \xi_i + f^{(2)}_{ij}\xi_i\xi_j + \cdots ,
\eeq
and the Grassmann integrals are defined as
\beq
	\int d\xi = 0~~,~~~ \int d\xi \xi = 1.
\eeq
Using these properties, the fermionic coherent state can be  
written explicitly as 
\beq
|\xi \rangle = \exp \Big\{-{1\over 2}\sum_i \xi^*_i \xi_i\Big\}
		\exp \Big\{ \xi_i c^\dagger_i\Big\} |0\rangle ,
\eeq
where $|0\rangle$ is the fermion vacuum state: 
$c_i|0\rangle=0$. For the 
fermionic coherent states, the resolution of 
identity can be written
similarly as
\beq
	\int \prod_i d\xi^*_i \xi_i |\xi \rangle \langle \xi | = I .
\eeq

Based on these properties of fermionic 
coherent states, the functional 
integral of fermion fields is rather 
easy to derive \cite{Ohn78}.  
Consider a fermion field $\psi$ coupling 
with a scalar boson field 
$\phi$, the Lagrangian is
\beq
{\cal L}(\overline{\psi},\psi,\phi) = \overline{\psi}(i\not \! 
	\partial - m)\psi +{1\over 2}[(\partial \phi)^2 -m^2 \phi^2] 
	- g \phi \overline{\psi}\psi .
\eeq
Following the same procedure, one obtains the functional integral 
\beq
W(\overline{\zeta},\zeta,J) = \int [d\overline{\xi}][d\xi][d\phi] 
	\exp \Big\{\int d^4 x \Big[{\cal L}(\overline{\xi},\xi,\phi) 
+ \overline{\zeta}\xi + \overline{\xi}\zeta + J\phi \Big] \Big\},
\eeq
where the fermionic sources $\zeta^*$, 
$\zeta$ are also Grassmann numbers.  In fact, Schwinger
first introduced such a generating functional for 
fermion fields, in order to derive 
the fermion field Green's functions 
\cite{Sch51}.

These results can be extended to quantum electrodynamics and 
quantum chromodynamics, although the later will be more complicated
because of non-abelian gauge fields \cite{Lee73,Fed80}.
In most of text books, the discussions on coherent states in field 
theory are restricted usually in contents of the above formulation. 
One may derive from such a formulation almost everything 
about perturbative field theory, such as Feynman rules, the 
perturbation expansion, and renormalization analysis, etc. At this 
point, the functional integral of quantum field theory in terms of 
coherent states is actually nothing special. It is the standard 
formulation that one can also obtain from other methods. If the 
field theory can be treated perturbatively, one can always solve
the theory in one or other ways, based on the developments of 
field theory in the last fifty years.  The challenge in field 
theory we faced today (or in the past three decades since 
the theory of the strong interaction, namely quantum chromodynamics, 
was proposed) is in the nonperturbation section.
There is no a systematic approach in field theory that one can 
used to completely solve a nonperturbation problem, such as the
vacuum structure in non-abelian gauge theory, or bound state problem
in strongly coupling systems. In the next few sections, I will
try to illustrate some specific problems to see if the generalized
coherent states developed later can play some useful roles
to the nonperturbation field theory of strongly interacted systems.  

\section{Squeezed Coherent States and Quantum Fluctuations}

The first example I go to discuss is how one may use squeezed states 
to study the low energy quantum fluctuations in strong interaction
theory. Different from what is discussed in the previous section
where the content may be found in text books, here I should point 
out that the formulation presented in this section has actually not 
been completed yet, and more work remains for further investigations. 

The squeezed states are a generalization of photon 
coherent states. Aagin, the squeezed states first attracted the 
attention in quantum optics \cite{Yao76,Fab92}. In the early 
development, the principal potential applications of squeezed 
states are in the field of optical communications and ``quantum 
nondemolition experiments" designed for the detection of 
gravity waves \cite{Wal83}.  Later on, because of the ``capacity'' 
of treating quantum fluctuations, squeezed states have been 
used in various subjects, such as quantum measurement theory, 
quantum nonlinear dynamics, molecular dynamics, 
dissipative quantum mechanics as well as in quantum gravity
and condensed matter physics.

In quantum optics, the uncertainty principle places a damper on 
the enthusiasm with which quantum engineers approach the problem 
of coding and transmitting information by optical means.  
Specifically, the quantum noise inherent in light beams places 
a limit on the information capacity of an optical beam.	
Since the uncertainty principle is a statement about areas 
in phase space, noise levels in different quadratures are statements 
about intersections of uncertainty ellipses with these axes.  
Any procedure which can deform or squeeze the uncertainty 
circle to an ellipse can in principle be used for noise 
reduction in one of the quadratures.  Such squeezing does 
not violate the uncertainty principle; rather, it places 
the larger uncertainty in a quadrature not involved in the 
information transmission process. A typical procedure for 
squeezing the error ellipse involves applying a 
classical source to drive two photon emission and absorption 
processes in much the same way that single photon processes 
can be used to generate a coherent state of the electromagnetic 
field.  

For simplification, one may consider a basic Hamiltonian 
describing two photon processes in a single mode 		
\beq
	H = \omega \Big(a^\dagger a + {1 \over 2}\Big) 
		+ f(t) a^{\dagger 2} + f^*(t) a^2 . \label{3sh}
\eeq
Then, the squeezed state can be obtained by 
directly solving the time dependent Schr\"{o}dinger equation.
%\beq
%	i\hbar {\partial \over \partial t} | \Psi \rangle
%		= H_s | \Psi \rangle .  \label{3shs}
%\eeq
If the initial state is the photon vacuum, a general 
solution is
\beq
	| \beta \rangle =  \exp\Big( {1\over 2}\beta
		a^{\dagger 2} - {1\over 2} \beta^*
		a^2 \Big) | 0 \rangle e^{i\varphi}. \label{3sqz}
\eeq 
This is the squeezed states generated by Eq.(\ref{3sh}). If one  
defines the photon's position and momentum coordinates $(x,p)$
in terms of the creation and annihilation operators as in the 
previous section, one can then find that
\bea
	&& \Delta x^2 = \langle \beta | x^2 | \beta \rangle
		= {1 \over 2} \Big|{\rm cosh}|\beta| +
		{\beta \over |\beta|} 
		{\rm sinh}|\beta|\Big|^2, \nonumber \\	
	&& \Delta p^2 = \langle \beta | p^2 | \beta \rangle = 
		{1 \over 2} \Big|{\rm cosh}|\beta| - {\beta 
		\over |\beta|} {\rm sinh}|\beta| \Big|^2,
		\label{uncty}
\eea
and $\Delta x \neq \Delta p$ but $\Delta x \Delta p \geq 
{1\over 2}$ (here we set $\hbar=1$). While the vacuum has a circle 
uncertainty ($\Delta x = \Delta p$) in phase space. 
This shows that the operator $D_{\rm sq}(\beta)=\exp\Big( 
{1\over 2}\beta a^{\dagger 2} - {1\over 2} \beta^* a^2 \Big)$
squeezes the uncertainty circle of a wave packet into an ellipes
so that quantum fluctuation (noise) can be reduced in one of 
the quadratures. 

In general, it is desirable to squeeze a field coherent state
which can be generated by		
\beq
	H = \omega (a^\dagger a + {1\over 2}) 
		+ f_2(t) a^{\dagger 2} +  + f^*_2(t) a^2
		+ f_1(t) a^{\dagger} + f^*_1 (t) a . \label{3scsh}
\eeq
The sequence in which the processes of coherent state 
formation and squeezing occur is governed by the time 
dependence of the functions $f_2(t)$ and $f_1(t)$.  
The general form of the state at the time $t$ can be 
expressed (apart from a phase factor) by
\beq
	| z\beta \rangle =  \exp \Big( z a^\dagger 
		- z^* a\Big)\exp \Big({\beta\over 2}
		a^{\dagger 2} - {\beta^* \over 2} a^2\Big) 
		| 0 \rangle , 	\label{3scs1} 
\eeq
where the complex variables $z$ and $\beta$ are functions
of the time $t$ in general. Eq.~(\ref{3scs1}) is usually 
called squeezed coherent state.
The physical process of the squeezed coherent states 
can be understood as follows: by first squeezing the vacuum 
(the wave packet) by the two photon excitations, and then 
displacing it as a photon coherent state by the external 
source.

Using group theory, one can show that the squeezed 
coherent states must also form a overcomplete set of states.
\beq
	\int {dz dz^*\over 2\pi} {df dg\over 2\pi} | z\beta 
		\rangle  \langle z\beta| = I
\eeq
where the variables $f$ and $g$ are introduced by Jackiw and 
Kerman \cite{Jac79} to characterize quantum fluctuations 
(noise) of the position and momentum:
\beq
	\Delta x^2 = f~,~~ \Delta p^2 = {1\over 4f} + 4f g^2 ,
\eeq
These two variables relate to the squeezing parameter 
$\beta$ by (\ref{uncty}). By the completeness, one can also
derive a path integral of quantum
mechanics in the squeezed coherent state representation:
\bea
	\langle z'(t_f)\beta'(t_f) | z(t_0)\beta(t_0) \rangle 
		&=& \int [dx(t)][{dp(t)\over 2\pi}][df(t)][{dg(t)
		\over 2\pi}] \nonumber \\
		& & ~~~~~ \times \exp \Big\{ i \int_{t_0}^{t_f} dt 
		\Big[{1\over 2}(p\dot{x}-x\dot{p}) - f\dot{g}-
		{\cal H}_{\rm eff}\Big] \Big\},
\eea
where the effective Hamiltonian is the matrix element of the 
Hamiltonian operator $[H=p^2/2 + V(x)$] in the squeezed coherent 
state:
\beq
{\cal H}_{\rm eff}(x,p,f,g) = {1\over 2}p^2 + {1\over 8f} + 2fg^2 
	+ \exp\Big({f\over 2}{\partial^2 \over \partial x^2}\Big)V(x).
\eeq
The expression of path integral in squeezed coherent states shows
that $f$ and $g$ which characterize quantum fluctuations 
become a pair of conjugate variables.
The extremal values of the exponent in the path integral leads
to the following generalized equations of motion:
\bea
	{dx\over dt}&=&\partial {\cal H}_{\rm eff}/\partial p~,~~ 
{dp\over dt}=-\partial {\cal H}_{\rm eff}/\partial x~, \nonumber \\
	{df\over dt}&=& \partial {\cal H}_{\rm eff}/\partial g~,~~ 
	{dg\over dt}=-\partial {\cal H}_{\rm eff}/\partial f~.~~ 
\eea
Physically, the equations of motion for $(x,p)$ determine the time 
evolution of the center of wave packets and those for $f,g$ 
characterize the time evolution of the quantum fluctuations 
(quadratures). Therefore, the variables 
($f,g$) describe the squeezing and spreading of quadratures in times, 
which provides a classical-like dynamical theory for the controlling 
of quantum noise and signal. 

It is worth pointing out that although the concept of squeezed 
state first attracted the attention in quantum optics, squeezed 
state itself was introduced much earlier by Valatin and Bulter in
the study of superfluidity \cite{Val58,Cum66}. 
Similar to the BCS state of superconductivity (which I 
will discuss later), Valatin's superfluid 
ground state is defined as
\bea
|\{z_k\beta_k\}\rangle &=& \exp \Big\{\sum_k (z_k a^\dagger_k
	-z^*_k a_k)\Big\} \exp \Big\{ \sum_k {1\over 2}(\beta_k 
	a^\dagger_k a^\dagger_{-k}-\beta^*_k a_k a_{-k}) \Big\} 
	|0\rangle \nonumber \\
	&=& \exp \Big\{\sum_k (z_k a^\dagger_k-z^*_k a_k)\Big\} 
		|\{\beta_k\}\rangle
\eea
which is the standard form of squeezed coherent states one currently 
used. In many-body picture, such states have two consequences. 
The squeezed operator $\exp \Big\{ \sum_k 
{1\over 2}(\beta_k a^\dagger_k a^\dagger_{-k} -\beta^*_k a_k 
a_{-k}) \Big\}$ acting on the trivial vacuum generates a canonical 
transformation of quasiparticles:
\beq
	\alpha_k = {\rm cosh} |\beta_k|a_k - {\beta_k\over |\beta_k|}
		{\rm sinh}|\beta_k| a^\dagger_{-k}~,~~
		\alpha_k  |\{\beta_k\}\rangle = 0.
\eeq  
Then using the quasiparticle vacuum to generate a bosonic 
coherent state. With such a state, one may develop a 
microscopic theory of superfluid helium, in which the 
normal and superfluid states become a direct analogy 
of noise and signal in partially coherent radiation fields.

We now use the squeezed coherent state to formulate a 
possible theory that may be useful in addressing the 
low energy quantum fluctuations in field theory.  Let us
 consider again the neutral scalar field ($\phi^4$ 
theory) as an example. One can define the squeezed 
coherent state of the field $\phi$ as \cite{Tsu91}
\bea
|\Psi \rangle &=& {\cal N}\exp \Big\{i\int d^3 x 
[\pi(x)\phi_{op}(x)- \phi(x)\pi_{op}(x)] \Big\} \nonumber \\
		& & ~~~~\times \exp \Big\{ \int d^3 x 
	d^3 y [\phi_{op}(x) D(x,y)\phi_{op}(y) \Big\}|0\rangle
\eea
where ${\cal N}$ is a normalization constant. This squeezed 
coherent state is also defined at a given instant time so 
that $t=t_x=t_y$. One can show that:
\bea
& & \langle \Psi | \phi_{op}({\bf x}) | \Psi \rangle = \phi(x) ~~,~~~
\langle \Psi | \pi_{op}(x) | \Psi \rangle = \pi(x) , \nonumber\\
& & ~~~ \langle \Psi | \phi_{op}(x)\phi_{op}(y)| \Psi \rangle = 
		\phi(x)\phi(y) + \Phi(x,y) , \nonumber\\
	& & \langle \Psi | \pi_{op}(x)\pi_{op}(y)| \Psi \rangle = 
		\pi(x)\pi(y) + {1\over 4}\Phi^{-1}(x,y) \nonumber \\
		& & ~~~~~~~~~~~~~~~~~~~~~~~~~~ + 4 \int d^3 x' d^3 y' 
		\Pi(x,x') \Phi(x',y')\Pi(y',y) ,
\eea	
where
\bea
	& & D(x,y) \equiv {1\over 2}[\Phi_0^{-1}(x,y)-\Phi^{-1}(x,y)]
		+ 2i \Pi(x,y), \nonumber \\
	& & ~~~~~ \Phi_0(x,y) = \langle 0| \phi_{op}(x)
		\phi_{op}(y)| 0 \rangle .
\eea
The squeezed coherent states of bosonic field are also overcomplete, 
namely
\beq
	\int [d\phi(x)][{d\pi(x)\over 2\pi}][d\Phi(x,y)][{d\Pi(x,y) 
		\over 2\pi}] |\Psi \rangle \langle \Psi | = I .
\eeq 
Following the similar procedure discussed in the previous
section, one can derive the functional integral $W(J)$ in the
squeezed coherent state representation:
\bea
 	W(J) &=& \int [d\phi(x)][{d\pi(x)\over 2\pi}][d\Phi(x,y)]
		[{d\Pi(x,y)\over 2\pi}] \nonumber \\
		& & ~~~~\times \exp \Big\{\int d^4 x \Big[ {1\over 2}
		(\pi \dot{\phi}-\phi \dot{\pi}) -\Phi \dot{\Pi} 
	- {\cal H}_{\rm eff} (x)] + J(x)\phi(x)\Big] \Big\}, 
		\label{ssfi}
\eea
where
\bea
{\cal H}_{\rm eff} &=& {1\over 2}[\pi^2(x) + (\bigtriangledown
	\phi)^2 +m^2\phi^2] +\exp\Big({\Delta(x) \over 2}{\partial^2
		\over \partial\phi^2}\Big)V(\phi) \nonumber \\
& & ~~ + {1\over 8}\Phi^{-1}(x,x) + 2 \int d^3 x' d^3 y' \Pi(x,x') 		\Phi(x',y')\Pi(y',x) \nonumber \\
		& & ~~+ {1\over 2}\Big[\lim_{x\rightarrow y}
		(\bigtriangledown_x \bigtriangledown_y \Phi(x,y)) 
		+m^2\Phi(x,x)\Big]
\eea
and $\Delta(x,t)=\lim_{x \rightarrow y} [\Phi(x,y)-\Phi_0(x,y)]$.
Here I have not integrated out the conjugate momentum $\{
\pi(x),\Pi(x)\}$ to obtain a covariant functional integral.
The physical picture of this new formulation is that besides 
the original field variable $\phi(x),\pi(x)$ in the Hamiltonian
formulation, quantum fluctuations, characterized by $\Phi(x,y)$
and $\Pi(x,y)$, are introduced as new dynamical field variables. 
These new dynamical field variables describe the low energy 
excitations (.i.e. the {\bf composite particles}) of  
strongly interacted systems. Similarly, taking the extreme 
value of the exponent in (\ref{ssfi}), one can find the equation 
of motion that determine the classical-like solution $\phi_0$:
\beq
	(\Box -m^2)\phi_0 + \exp\Big({\Delta(x) \over 2}{\partial^2
		\over \partial\phi_0^2}\Big) V'(\phi_0) = 0 ,
\eeq
which is coupled with the composite field $\Phi$.
The equation of motion for its quantum fluctuations by $\Phi$
is much more complicated.

Usefulness of the squeezed state functional integral is 
that one can derive an effective theory for the low energy 
composite particle fields coupling with the original fields. 
Here I may propose a procedure how to develop such an effective 
theory. First, one can determine the ``classical'' ground state by 
minimizing the effective Hamiltonian with respect to variables 
$\phi, \pi$ as well as $\Phi, \Pi$, which results in $\phi_0, (\pi_0=0)$
and $\Phi_0, (\Pi_0=0)$. Then, expanding the effective Lagrangian
near $(\phi_0, \pi_0, \Phi_0,\Pi_0)$ up to the second-order, 
namely only keeping the quadratic terms in $(\delta\phi,\delta 
\pi, \delta\Phi, \delta \Pi)$. Quantum effects become the 
time dependent fluctuations about the classical ground states. 
The corresponding linearized equations of motion determine the 
dispersions of quasiparticles and composite particles, denoted 
by $\omega_k$ and $\gamma_k$, respectively. For strong interaction 
field, usually $\omega_k > \gamma_k$ (due to the spontaneously 
symmetry breaking). Thus, in the low energy scale $(\omega_k
> \mu \geq \gamma_k)$, the composite particles and the 
quasiparticles are decoupled. Only the composite particles are 
kept with the high order corrections (as a perturbation) to 
form the low energy effective Lagrangian.  In the intermediate
energy scale $(\mu \sim \omega_k)$, both the composite particles 
and quasiparticles become active and coupled each other. 
The effective theory is then determined by the Lagrangian of the
composite particles coupled with quasiparticle degrees of freedom. 
In a rather high energy scale ($\mu >> \omega_k$), $\Phi$ 
and $\Pi$ should spread averagely over the entire space-time 
space such that only the original field variables remain. 
Thus, the theory returns back to the original one in high 
energy region. 

The above procedure is different from the conventional 
procedure of constructing a low energy effective theory. 
In the conventional approach, the low energy effective 
theory is constructed by separating the field variables into
the low energy and high energy parts. Then, using the functional
integral discussed in the previous section to integrate out the high 
energy part. The resulting Lagrangian is an effective Lagrangian
for the low energy physics. The advantage of the conventional
approach is that one can use the powerful renormalization group 
analysis of Wilson \cite{Wil74} to extract universal scaling 
properties contained in the theory, without explicitly solving 
the theory itself. However, in reality, physical
degrees of freedoms must also be very different in different energy 
scales. A typical example is the strong interaction in which
the degrees of freedom are quarks and gluons in high energy
region. But in low energy region, the degrees of freedom 
become hadrons which are composite particles of quarks and gluons.
Thus, in the conventional approach, the effective theory 
cannot catch the right physical degrees of freedom. Therefore, 
beyond the critical phenomena, the conventional approach may have
its limitation in applications. The squeezed coherent state 
formulation of the functional integral may provide 
a new method for the developments of effective field theory, 
although at this point a lot of work remains for further 
investigations. 
 
A potential application of the squeezed coherent states in 
field theory is the Yang-Mills gauge theory, especially
the color SU(3) gauge theory in quantum chromodynamics. Of 
course, the situation in quantum chromodynamics 
is much more complicated. Because
of the nonlinear properties in non-abelian gauge theory, the
conventional functional integral is already quite complicated.
However, the conventional functional integral in non-abelian
gauge theory is only useful for the derivation of covariant 
Feynman rules and the analysis of renormaliability. In other
words, it is only useful for perturbation calculations. 
As it is well-known, the difficulty of QCD lies in its 
nonperturbation domain, where quantum fluctuation must be strong.
Furthermore, the field strength of non-abelian gauge contains
the single as well as double gauge boson emissions and absorptions:
\beq
	F^a_{\mu \nu} = \partial_\mu A^a_\nu - \partial_\nu A^a_\mu
		- g f^{abc}A_\mu^b A_\nu^c .
\eeq
The squeezed coherent states may be the natural quantum states
 describing non-abelian gauge fields. 
If one can complete and extend the above formulation to the 
non-abelian gauge theory, then it may be able to develop 
a low energy effective theory for non-abelian gauge fields. 
I believe that such a low energy effective theory should be 
capable in dealing with gluon condensation, gluball states 
as well as low energy gluon dynamics. 

\section{Spin Coherent States and Non-Linear Sigma Model}

So far, I have only discussed bosonic-type coherent states in field
theory. However, coherent states, in terms of the language of 
group theory, are embedded in a topologically nontrivial geometrical 
space which involve a deep implication in physics \cite{Zha90}. 
The simplest coherent state carried a topologically
nontrivial geometrical space is the {\it spin coherent state}.  
The most attractive property in spin coherent states is that 
its topological structure naturally induces Dirac's {\bf magnetic 
monopole} \cite{Yan75}. Meanwhile, 
the spin coherent state representation 
of the path integral for a multi-spin system gives a realistic 
realization of Non-Linear Sigma Model which is an important
field theory model in condensed matter physics and particle
physics. Thus, before I go to discuss the general physics
implication containing in the geometrical structure of coherent
states, it may be useful to illustrate first the spin coherent 
state in details.

Let us start with a simple example: a spin-$1/2$ particle 
in a varying magnetic field: $B(t) = 
(B_x(t), B_y(t), B_z(t))$, described by the Hamiltonian,  
\beq
	H(t) = - \mu S \cdot B(t).	\label{1sh}
\eeq
Here $\mu$ is the particle's magnetic moment,
$\vec{S} = ( S_x,S_y,S_z)$ the spin operator that 
satisfies the usual angular momentum commutation relationship.
%\beq
%	[S_x, S_y] = i S_z~,~~ [S_y, S_z] = iS_x~,~~ [S_z,S_x]=iS_y.
%\eeq
The evolution of system governed by (\ref{1sh}) can be determined
by the Schr\"{o}dinger equation, whose general solution can be 
written as
\beq
	| \psi(t)\rangle = \alpha(t) | \downarrow \rangle + \beta(t)
		| \uparrow \rangle.	\label{1ss}
\eeq
Substituting (\ref{1ss}) into Schr\"{o}dinger equation, it is easy 
to determine the time dependence of the parameters $\alpha(t)$ 
and $\beta(t)$. However, in order to derive the spin coherent 
state, here, I may only concentrate on the structure of the 
state (\ref{1ss}).  The normalization of (\ref{1ss}) results 
in a constraint on the parameters $\alpha(t)$ and $\beta(t)$:
\beq
	| \alpha(t)|^2 + | \beta(t)|^2 = 1.
\eeq
If I parameterize $\beta=\sin{\theta\over 2}e^{-i\varphi}$, 
Eq.~(\ref{1ss}) can be expressed as
\beq
	| \psi (t) \rangle =  \Big(\cos {\theta(t) \over 2} 
		+ \sin {\theta(t) \over 2}
		e^{-i\varphi(t)} S^+\Big)|\downarrow \rangle 
		e^{i\phi(t)} ,	\label{1ss1}
\eeq
the raising and lowing spin operators $S^{\pm}$ are defined by
$S^{\pm} = S_x \pm i S_y$, and $S^+ |\downarrow \rangle =  
|\uparrow\rangle$ , $S^- |\uparrow \rangle = |\downarrow \rangle$.
Furthermore, one can easily show that
\beq
	 \Big(\cos {\theta \over 2} + \sin {\theta \over 2} 
		e^{-i\varphi} S^+\Big)| \downarrow \rangle =
		\exp\Big\{ {\theta \over 2} e^{-i\varphi} S^+ 
		- {\theta \over 2}e^{i\varphi} S^- \Big\} | 
		\downarrow \rangle \equiv | \theta \varphi \rangle.	
		\label{1.8}
\eeq
Then, (\ref{1ss1}) can be simply expressed as
\beq
| \psi (t) \rangle = | \theta(t) \varphi(t) \rangle e^{i\phi(t)}~. 
	 \label{1cssp}
\eeq
%and $\Omega (\eta) =  \exp\Big\{ \eta S^+ - \eta^* S^- \Big\}$.
The state  $|\theta \varphi \rangle $ is a standard 
expression of the spin coherent states.

As I mentioned a very important property of the spin coherent state 
is that $|\theta \varphi\rangle$ is embedded in a topologically 
nontrivial geometrical space, i.e., a two-dimensional sphere $S^2$. 
This can be varified directly from (\ref{1ss1}). Since $S_i = 
\sigma_i/2$ where $\sigma_i$ is Pauli matrix, 
\bea
	D_s(\eta)&\equiv&\exp\Big\{ \eta S^+ - \eta^* S^- \Big\} 
		= \exp \left[\begin{array}{cc} 0 & \eta \\ 
		-\eta^* & 0 \end{array} \right]  \nonumber \\
	&=& \left[
		\begin{array}{cc} \cos |\eta| & {\eta \over |\eta|}
		\sin | \eta| \\  -{\eta^* \over |\eta|} \sin|\eta|
 		& \cos |\eta| \end{array} \right] = \left[
	\begin{array}{cc} x_0 & x \\ -x^* & x_0 \end{array} \right]
		\label{1scm}
\eea
and $x_0$ is real while $x= x_1 + i x_2$.  Also, $D_s(\eta)$ 
is a unitary operator which leads to
\beq
	x_0^2 + |x|^2 = x_0^2 + x_1^2 + x_2^2 = 1.
\eeq
In other words, the parameter space of $D_s(\eta)$ is a
two-sphere $S^2$ with $\eta={\theta \over 2} 
e^{-i\varphi}$. Therefore, the spin coherent states are 
ono-to-one corresponding to the points on $S^2$ except for 
the north pole where it is ambiguous since all values of $\varphi$ 
correspond to the same point.

However, in defining the above spin coherent states, there is 
an ambiguous. For example, one can also define the spin
coherent states as 
\beq
	|\theta \varphi \rangle' = \Big(\cos {\theta \over 2}
		e^{i\varphi} + \sin {\theta \over 2}  S^+\Big)|
		\downarrow \rangle ,	\label{1.8a}
\eeq
which are also ono-to-one corresponding to the points in $S^2$ but
except for the south pole. These two spin coherent states are
related simply by a phase factor, 
\beq
	 |\theta \varphi\rangle'=e^{i\varphi}|\theta \varphi\rangle .
\eeq 
Geometrically, these two coherent states define the two ``patches'' of 
$S^2$. Since these two states are only different by a phase factor,
quantum mechanically, they must be equivalent. This implies that there 
is a gauge degree of freedom in the spin coherent states. 

To see clearly the physical implication induced by the topological 
structure of spin coherent states, one can construct the path 
integral of a quantum spin system. The spin coherent state also
obeys the overcompleteness:
\beq
	\int d\mu(\theta\varphi)
		|\theta \varphi \rangle \langle \theta \varphi |=I,
\eeq
where $d\mu(\theta\varphi)=\sin \theta d\theta d\varphi/2\pi$
is an invariant measure on $S^2$. Then, it is easy to show that the 
path integral of quantum mechanics for $H=H(\vec{S})$ is given by
\beq
\langle \theta'(t_f) \varphi'(t_f)|\theta(t_0) \varphi(t_0) \rangle 
		= \int [d\mu(\theta\varphi)] \exp\Big\{i \int_{t_0}^{t_f} 
		dt \Big[\langle \theta \varphi|i {d \over dt} |\theta 
	\varphi \rangle - \langle \theta \varphi|H(\vec{S})|\theta 
		\varphi \rangle \Big]\Big\} .
\eeq	 
In this path integral, the first term in the exponent, 
\beq
\omega[\theta\varphi] \equiv \int _{t_0}^{t_f} dt \langle \theta 
		\varphi|i {d \over dt }|\theta \varphi \rangle
		= \int_{\varphi_0}^{\varphi_f} {1\over 2}(1-\cos 
		\theta)d \varphi ,  \label{ssp1}
\eeq
is pure geometric that only depends on the trajectory 
over the sphere, but not on its explicit time dependence. 
For a closed path, $\omega[\theta\varphi]$ is actually a 
{\bf Berry phase} of the spin 
history \cite{Ber83}. Therefore, $\omega[\theta\varphi]$ is a gauge 
invariant one-form defined on the sphere $S^2$:
\beq
	\omega[\theta\varphi] = \int_{\varphi_0}^{\varphi_f} A_\varphi
		d\varphi = \int_{\varphi_0}^{\varphi_f} {\cal A}({\bf n})
		d{\bf n} ,  \label{ssp2}
\eeq
where ${\bf n}=(\sin\theta\cos\varphi,
\sin\theta\sin\varphi, \cos\theta)$ is a unit vector, 
and ${\cal A}({\bf n})$ is a unit vector potential. 
Compare with (\ref{ssp1}) and (\ref{ssp2}), one can find that
\beq
	{\cal A}^a = {1-\cos \theta \over 2 \sin \theta} \hat{\varphi} .
\eeq
This vector potential has one singularity at the south pole. 
It is this singularity where 
the Dirac string which carries the magnetic monopole flux enters 
the sphere. Hence, ${\cal A}^a$ is nothing but a gauge potential 
of Dirac's magnetic monopole.  Similiarly, for the spin coherent 
state $|\theta \varphi \rangle'$, the corresponding gauge potential is 
\beq
	{\cal A}^b = {1+\cos \theta \over 2\sin \theta} \hat{\varphi} .
\eeq
${\cal A}^a$ and ${\cal A}^b$ define the two non-singular patches 
of the monopole section. Their difference is a pure U(1) gauge 
in the overlapping equatorial region, $S^1$,
\beq
	{\cal A}^b_\varphi={\cal A}^a_\varphi + d\varphi 
		={\cal A}^a_\varphi -i g^{-1}dg.
\eeq
where $g=e^{i\varphi} \in $ U(1).
%or
%\beq
%	dW={\bf A}^a \cdot d{\bf n} - {\bf A}^b \cdot d{\bf n}
%		= {\bf n} \times d{\bf n} .
%\eeq

The existence of the above gauge degrees of freedom can 
be understand clearly by looking at the general 
definition of coherent states based on group theory 
\cite{Per72,Gil72}. In group theory, quantum states of a 
spin system form a unitary representation $V^s$ of the SU(2) group, 
here $s$ is an arbitrary spin. Choosing a fixed state, such as the 
lowest-weight state $|ss_z\rangle=|s-s\rangle \in V^s$, 
one can define spin coherent states as 
\beq
	|g \rangle_s = g |s-s\rangle~, ~~g \in SU(2) .
\eeq
In general, $g = \exp (i\alpha S_x)\exp( i \beta S_y)\exp(i\gamma
S_z)=\exp ({\theta \over 2} e^{-i\varphi} S^+ - {\theta \over 2} 
e^{i\varphi} S^-)\exp(i\gamma' S_z)$. Note that this 
decomposition, called the Baker-Campbell-Hausdorff formula, is unique. 
As a result, one can rewrite the above spin coherent states as
\beq   
	|g \rangle_s = \exp ({\theta \over 2} e^{-i\varphi} S^+ - 
		{\theta \over 2} e^{i\varphi}S^-) \exp (i\gamma' S_z)
		|s, - s \rangle =|\theta \varphi \rangle e^{i\chi}.\,
\eeq
where 
\beq
	|\theta\varphi \rangle = \exp ({\theta \over 2} e^{-i\varphi}
		 S^+ - {\theta \over 2} e^{i\varphi} S^-)|s,-s\rangle
\eeq
is the standard definition of spin coherent state for an arbitrary 
spin $s$ \cite{Are72}. Spin $s=1/2$ discussed above is a special 
case. As one can see, apart 
from a phase factor, the spin coherent states
can be generated by a unitary spin rotational operator acting on 
the fixed state $|s-s\rangle$. The unitary operator $\exp ({\theta 
\over 2} e^{-i\varphi} S^+ - {\theta \over 2} e^{i\varphi} S^-)$ 
is the coset representation of the space $SU(2)/U(1) 
\simeq S^2$. Therefore the sphere $S^2$ determines the topological 
structure of spin coherent states.  The magnetic monopole 
potential ${\cal A}^a$ defines a $U(1)$ fibre bundle over 
this sphere $S^2$. Meanwhile, the spin coherent states contain an
arbitrary phase $\chi$. In quantum mechanics, a quantum state is 
specified up to a phase factor, namely, physics is invariant 
for different choices of such a phase factor so that this phase 
factor is usually ignored.  However, when quantum states are 
embedded in a topologically nontrivial space, this phase freedom 
is indeed the associated gauge degrees of freedom of the fibre 
bundle over the space. In spin coherent states, 
the phase $\chi$ is just the gauge degree of freedom that connects 
different choices of magnetic monopole potentials. Ignoring 
(or fixing) this phase factor corresponds to a gauge fixing.

Furthermore, the topological properties of the spin coherent 
states also play an important role in the study of spin dynamics. 
A typical example is the Heisenberg model in condensed matter 
physics. Heisenberg model is used to 
understand quantum magnetism of strongly correlated
electron systems. The model Hamiltonian considered here is very 
simple:
\beq
	H_{J} = J \sum_{\langle i,j\rangle} {\bf S}_i \cdot {\bf S}_j
\eeq
which describes a many-spin (each spin $=s$) system with the nearest 
neighbor exchange interaction. ``Classically'', the ground state 
of the above Hamiltonian is easily determined. When $J>0$, the 
minimum energy is given by the state in which the nearest-neighbor 
spins are always anti-alignment. These states are called in 
literatures the N\'{e}el states.  Correspondingly, the system is 
an antiferromagnet. If $J<0$, the ground state is simply given
by the state with all spin aligned in the same direction, which
is a ferromagnetic state. These consequences can be obtained
explicitly by taking the spin coherent state
\beq	\label{scshm}
	|\{\theta_i\varphi_i\} \rangle = \prod_i \exp \Big\{{\theta_i 
		\over 2} e^{-i\varphi_i}S_i^+ -{\theta_i 
		\over 2} e^{i\varphi_i} S_i^-\Big\} |s-s\rangle
\eeq
as a trial wave function and minimizing the model Hamiltonian
\beq
	\delta (\langle \{\theta_i\varphi_i\}| H_J
		|\{\theta_i\varphi_i\}\rangle) 
		= \delta(Js^2\sum_{\langle i,j\rangle}[\cos\theta_i 
	\cos\theta_j + \sin\theta_i \sin \theta_j \cos(\varphi_i
		-\varphi_j)])=0.
\eeq
The resulting ground state is given by
\beq  \label{cgs}
	\left\{ \begin{array}{l} J>0 \rightarrow \theta_{i+1}=
		\pi-\theta_i, \varphi_{i+1} = \varphi_i+\pi \rightarrow
	{\rm antiferromagnet} \\ ~~ \\ J<0 \rightarrow \theta_{i+1}=
		\theta_i, \varphi_{i+1} = \varphi_i \rightarrow
		{\rm ferromagnet} \end{array} \right. .
\eeq
An important concept one can obtain from the above result is that the
ground state spontaneously breaks the global spin rotational symmetry.  
As one can check the model Hamiltonian is invariant under global 
spin rotational transformations: $T=\exp(i\vec{\alpha}\cdot {\bf S})$, 
where ${\bf S}=\sum_i {\bf S}_i$. While the ground state energy 
does not change when all the spins in (\ref{cgs}) are globally
rotated. This leads to a SO(3) degeneracy of the ground states, 
namely these ground states have a lower symmetry than the Hamiltonian. 
Such a situation is called the {\bf spontaneously symmetry breaking}.  
Quantum mechanically, it leads to gapless spin-wave excitations
in the Heisenberg model, and Goldstone bosons in general. 

The quantum dynamics of interacting spins can be studied from 
the time-evolution of the system at zero-temperature. The 
time-evolution is determined by the Green's function which is defined 
by the matrix element of the evolution operator between two spin
coherent states:
\beq
	G(t_f,t_0) = \int \prod_i[d\mu(\theta_i\varphi_i)] \exp\Big\{i 
		{\cal S}[\theta_i(t), \varphi_i(t)] \Big\},
\eeq
here, the effective action is given by
\bea
	{\cal S}[\theta_i(t),\varphi_i(t)] &=& \int^{t_f}_{t_0} dt 
		\Big[\langle \{\theta_i \varphi_i\}|i {d \over dt} 
		|\{\theta_i \varphi_i\}\rangle - \langle \{\theta_i 
		\varphi_i\}|H_J|\{\theta_i \varphi_i\}\rangle \Big] 
		\nonumber \\
	&=& \int^{t_f}_{t_0} dt \Big[s\sum_i (1-\cos\theta_i)
		{d\varphi_i \over dt} \nonumber \\
	& &~~~ - J s^2 \sum_{\langle i,j\rangle}[\cos\theta_i 
		\cos\theta_j + \sin\theta_i \sin \theta_j \cos(\varphi_i
		-\varphi_j)]\Big]	\label{hsd}
\eea
Note that the thermal dynamics can be obtained in a similar
form. The partition function can be expressed in terms of 
a spin coherent state path integral as well:
\bea
{\cal Z}(\beta) &=& \langle \{\theta_i\varphi_i\}|\exp \Big\{-\beta
		H_J \Big\} |\{\theta_i\varphi_i\} \rangle \nonumber \\
		&=& \int [d\mu(
		\beta_i(\tau)\varphi_i(\tau))] \exp\Big\{-{\cal S}
		[\theta_i(\tau),\varphi_i(\tau)] \Big\}
\eea
with
\bea
	{\cal S}[\theta_i(\tau),\varphi_i(\tau)] &=& \int_0^\beta 
		d\tau \Big[-is\sum_i (1-\cos\theta_i){d\varphi_i \over 
		d\tau} \nonumber \\
	& &~~~ + J s^2 \sum_{\langle i,j\rangle}[\cos\theta_i 
		\cos\theta_j + \sin\theta_i \sin \theta_j \cos(\varphi_i
		-\varphi_j)]\Big]	, \label{hsd1}
\eea
where $\tau$ is an imaginary time from 0 to $\beta=1/kT$.

Eq.~(\ref{hsd}) shows that spin dynamics is induced by the geometrical
phase $\omega=\int dt \sum_i s(1-\cos\theta_i)
\dot{\varphi}_i$ which contains the time derivative and therefore
leads to the equation of motion for $\{\theta_i(t),\varphi_i(t)\}$. 
It also shows that the magnetic 
monopole potential $A_{\varphi_i}$ is actually the conjugate 
momentum of $\varphi_i$ in spin dynamics. If one defines the 
generalized position and momentum coordinates by
\beq
	q_i = \varphi_i~,~~ p_i = s(1-\cos \theta_i),
\eeq
and expands the effective action ${\cal S}[\theta_i(t),\varphi_i(t)]$
around the ground state $\{q_i^0, p_i^0$; $q_{i+1}^0, p_{i+1}^0\}$
[given by Eq.~(\ref{cgs})]:
\bea
	q_i = q_i^0 + \delta q_i~~&,&~~ p_i = p_i^0 + \delta p_i~~,
		\nonumber\\
	q_{i+1} = q_{i+1}^0 + \delta q_{i+1}~~&,&~~ p_{i+1} = p_{i+1}^0 
	+ \delta p_{i+1}~~,
\eea
up to the quadratic terms, one can determine explicitly the 
disperson of {\bf spin-wave excitations} \cite{Mat88,Wit87}. 

However, since $p_i$ is related to the magnetic monopole potential,
the conjugate coordinates $\{q_i,p_i\}$ used above are indeed 
gauge dependent quantities. To explore the 
possible topological effects in spin dynamics, it is better 
to use a gauge invariant formulation. This may be done by using a
global notation ${\bf n}$ (a unit vector) to represent the spin 
direction, without specifying the parameterization of the sphere 
by $\theta,\varphi$. Then, the Green's function can be expressed 
as:
\beq
	G(t_f, t_0) = \int [d\mu({\bf n}_i)] \exp\Big\{i 
	\int^{t_f}_{t_0} dt \Big[s\sum_i {\cal A}_i\cdot \dot{\bf n}_i
		- J s^2 \sum_{\langle i,j\rangle} {\bf n}_i \cdot
		{\bf n}_j \Big] \Big\}.
\eeq
Taking the continuum limit
\beq
	{\bf n}_i \rightarrow c_i {\bf n(x_i)},
\eeq
where $c_i = 1$ ($e^{i {\bf x}_i \cdot \vec{\pi}}$) for the ferromagnet 
(antiferromagnt), $|{\bf n(x_i)}|=1$, and $\vec{\pi}=(\pi, \cdots, \pi)$,
${\bf x_i} \in R^d$. Then the Green's function can be expressed as
\beq
	G(t_f,t_0) = \int [d\mu({\bf n(x)}] \exp \Big\{ i \int d^{d+1}
		{\cal L}({\bf n})\Big\},
\eeq
where ${\cal L}(\bf n)$ is an effective Lagrangian. In the low energy
(long wave-length) limit, it is reduced to the Lagrangian of the 
{\bf Non-Linear Sigma Model} in $d+1$-dimensional space \cite{Hal83},
\beq
	{\cal L}({\bf n}) = {1\over 2g} \partial_\mu {\bf n} \cdot 
		\partial^\mu {\bf n} + \cdots
\eeq
where ``$\cdots$'' dnotes the high order derivatives. This Lagrangian
ensures the existence of gapless spin-wave excitations. Such a 
Non-Linear Sigma Model has been widely studied in condensed matter
physics, including the problems in quantum magnetism, quantum Hall 
effect and disorder dynamics.

\section{Generalized Coherent States and Nonabelian Gauge Fields}

In this last section, I will discuss the generalized coherent states 
and their potential applications in field theory. Generalization of 
coherent states is based on group theory developed by Perelomov and
also Gilmore \cite{Per72,Gil72}. The spin coherent state discussed
in the previous section is an example of such generalization. 
Actually, Glauber had  
pointed out in his seminal paper \cite{Gla63} that the photon coherent 
states can be constructed starting from any one of three mathematical 
definitions. 
\begin{itemize}
 \item Definition 1.  The coherent states $| z \rangle$  
are eigenstates of the annihilation operator $a$:		
\beq
	a | z \rangle = z | z \rangle .	
\eeq
\item Definition 2. The coherent states $| z \rangle$ are 
quantum states with a minimum uncertainty relationship:		
\beq
	\Delta x^2 \Delta p^2 = {\hbar^2 \over 4}	\label{2unc1}
\eeq
\item Definition 3.  The coherent states $| z \rangle$ can
be obtained by applying a displacement operator $D(z)$ on the 
ground state of harmonic oscillator:		
\beq \label{gdcs}
	| z \rangle = D(z) | 0 \rangle~~,~~~
	D(z) = \exp(za^\dagger - z^*a) .	
\eeq
\end{itemize}
We have analyzed these definitions and pointed out [Zha90] that 
the generalization of eigenstates of the lowering operator 
is not always possible. Indeed, the adoption of this definition 
to generalize the coherent state concept has two major drawbacks: 
a).~Coherent 
states cannot be defined in Hilbert spaces of finite dimensionality 
in this way, as we have seen for the spin systems. b).~The states 
so defined do not correspond to physically realizable states, 
except under special circumstances that the commutator of the 
annihilation operator (or lowering step operator) and its hermitian 
adjoint is a multiple of the identity operator.  Therefore, under 
this condition one restricts oneself to the bosonic field. As a 
result, the generalization based on definition 1 to other dynamical 
systems is not always applicable.  

On the other hand, the generalization 
based on the definition 2 is by no means unique.  
The bosonic (photon) coherent states are the minimum uncertainty 
states essentially because they are non-spreading wave packets.   
Although the minimum uncertainty states are physically very 
interesting, the generalization along 
this direction has several limitations: a).~These coherent states 
can only be constructed for the classically integrable systems 
in which there exists a set of canonical coordinates and 
momenta such that the respective Hamiltonians can be reduced 
to quadrature.  This condition requires a flatness condition 
on the operator algebra which reduce the commutation relations 
to those of the standard bosonic creation and annihilation 
operators.  b).~The wave packets with the minimum uncertainty 
are not unique. Different ones may have different properties.  
Also, such states may be incomplete, or even if they are complete 
it is not certain that the standard form of a resolution of unity  
exists.  Thus, minimum uncertainty states appear to have 
few, if any, useful properties.     	

In literatures, the realization of generalized coherent 
states are indeed achieved based on displacement operators. 
The basic theme of this development was to intimately connect 
coherent states with dynamical symmetry groups of a physical 
problem.  Since all physical problems formulated 
in quantum theory have a dynamical group (although sometimes 
the group may be too large to be useful), an important outcome 
of this recognition is that coherent states can be generalized 
to all the quantum problems. 

I should outline here a generalization procedure how an arbitrary 
coherent state can be generated by displacement operators.
Consider a set of operators $\{ T_i \}$ closed under commutation:		
\beq
	[ T_i~, ~T_j] = T_iT_j - T_jT_i = \sum_k C^k_{ij} T_k , 
		\label{4als}
\eeq
That is, $\{ T_i \}$ span a algebra {\bf g}, and $C_{ij}$ 
in (\ref{4als}) are structure constants of {\bf g}.  If {\bf g} 
is a semisimple Lie algebra, it is more convenient to write  
$\{ T_i \}$ in terms of the standard Cartan basis $\{H_i, E_\alpha, 
E^\dagger_\alpha = E_{-\alpha} \}$:		
\bea
	&& [ H_i~ ,~ H_j] = 0~~, ~~~ [ H_i~, ~E_\alpha  ] = 
		\alpha_i E_\alpha , \nonumber \\
	&& [ E_\alpha , E_{-\alpha} ] = \alpha^i H_i ~~, ~~~
	[ E_\alpha~, ~E_\beta ] = N_{\alpha \beta} E_{\alpha+\beta}.
\eea
In quantum theory, for such a given set of closed operators 
$\{ T_i \}$, the quantum states are described by a Hilbert 
space $V^{\Lambda}$ which is a representation of {\bf g}. Let
{\bf G} be the covering group of {\bf g}. The Hilbert 
space $V^{\Lambda}$ carries a unitary irreducible representation 
$\Gamma^{\Lambda}$ of {\bf G}.  One may choose a normalized 
state $|\phi_0 \rangle$ in the Hilbert space $V^{\Lambda}$ 
as a fixed state. Then the generalized coherent state is generated
by an element $g\in {\bf G}$ acting on the fixed state 
$|\phi_0\rangle$.
\beq
	|g\rangle_{\bf G} = g | \phi_0 \rangle.
\eeq
In group theory, every element $g \in$ {\bf G} can be
uniquely decomposited into a product of two group elements:
$g = k h$, here one should require $h \in$ {\bf H} such that 		
\beq
	h  | \phi_0 \rangle =  | \phi_0 \rangle e^{i\chi} ,
\eeq
and {\bf H} is the maximum subgroup of {\bf G} that 
leaves the fixed state invariant up to a phase factor. 
While $k$ is an operator of the coset space {\bf G/H}.
If {\bf G} is a semisimple Lie group and $|\phi_0\rangle$ 
is the lowest weight state, $k$ can be generally 
written as 
\beq
	k \equiv D_G(\eta)=\exp\Big\{ \sum_{\alpha>0}(\eta_\alpha 
		E_\alpha - \eta^* E_{-\alpha})\Big\} \in {\bf G/H}.
\eeq
This operator $D_G(\eta)$ is usually called a displacement operator of 
{\bf G}, which gives a coset representation of {\bf G/H}. 
As a result, 
\beq
	|g \rangle_{\bf G}  = D_G(\eta) | \phi_0 \rangle e^{i\chi}
		= | \Phi(Z) \rangle e^{i\chi},	\label{gtcs}
\eeq
Perelomov and Gilmore \cite{Per72,Gil72} define the state $|\Phi(Z)
\rangle$ as the generalized coherent states of {\bf G}:
\beq
	|\Phi(Z)\rangle = D_G(\eta)| \phi_0 \rangle ={\cal N}(Z) 
		\exp \Big\{ \sum_{\alpha>0} Z_\alpha E_\alpha \Big\}
		|\phi_0 \rangle ,
\eeq
and ${\cal N}(Z)$ is a normalized constant.
The generalized coherent states defined in such a way have two 
important properties 
\begin{itemize}
\item The set of the generalized coherent states satisfies:
\beq	\label{1com}
	\int d\mu(Z) |\Phi(Z)\rangle \langle \Phi(Z)| = I,
\eeq
where $d\mu(Z)$ is the {\bf G}-invariant Haar measure on {\bf G/H}.
\item The generalized coherent states are one-to-one corresponding
to the points in the coset space ${\bf G/H}$ except for some singular
points, such as the north pole or south pole of the two-sphere in 
spin coherent states. Therefore, the generalized coherent states
are embedded into a topologically nontrivial geometrical space.
\end{itemize}

Systems discussed in the previous sections are only some simple
examples of the generalized coherent states. 
The harmonic oscillator admits a dynamical group 
$H_4$, called Heisenberg-Weyl group. The photon 
coherent states are obtained via a one-to-one correspondence 
with the geometrical coset space H(4)/U(1)$\times$U(1) (a complex
plane) by the displacement operator $D(z)\in$ H(4)/U(1)$\times$U(1). 
The two-photon processes has a SU(1,1) dynamical group. The 
squeezed states are obtained by the displacement (squeezed) 
operator $D_{\rm sq}(\beta)\in$ SU(1,1)/U(1) (a two-dimensional 
hyperboloid space). And the spin coherent states discussed in 
the previous section are generated by the displacement operator 
$D_s(\theta\varphi)\in$ SU(2)/U(1) (a two-dimensional sphere). 

I should emphasize here that the phase $\chi$ in the group-theoretical
coherent state (\ref{gtcs}) is the {\bf H}-gauge degrees of freedom 
over the coset space {\bf G/H}.  All the three sets of coherent 
states discussed in the previous sections contain an U(1) gauge, 
but only the sphere (spin) carries a nontrivial fibre bundle
so that the gauge degrees of freedom become important. 
To obtain a non-abelian gauge, one must consider the 
generalized coherent states of a group {\bf G} whose rank is 
larger than one such that {\bf H} can be a non-abelian group.
  
To examine non-abelian gauge degrees of freedom in the generalized 
coherent states, one may extend the path integral to the generalized 
coherent state representation.  The Green's function is now 
defined as the matrix element of the evolution 
operator in the generalized coherent states:
\beq
	G(t_f, t_0) = \langle \Phi'(Z)| T \exp \Big\{ - i 
		\int_{t_0}^{t_f} H(t) dt \Big\} | \Phi(Z) \rangle .
\eeq
Following the same procedure as it has been done in the previous 
sections that divides the time interval $t_f-t_0$ into 
$N$ intervals, each with $\varepsilon = (t_f-t_0)/N$, 
then inserts the resolution of identity (\ref{1com}) at each 
interval point, and finally lets $N$ go to infinity, the 
Green's function can be expressed as a generalized coherent state
path integral. 
\bea
	G(t_f,t_0)&=& \lim_{N \rightarrow \infty} \int 
		\Big(\prod_{i=1}^{N-1} d\mu_i(Z)\Big) \prod_{i=1}^N 
		\langle \Phi_i(Z)|\exp\Big\{-i\varepsilon H(t_i) \Big\} | 
		\Phi_{i-1}(Z) \rangle \nonumber \\
%	&=& \lim_{N \rightarrow \infty} \int \Big( \prod_{i=1}^{N-1} 
%		d\mu_i (Z)\Big) \exp \Big\{i \sum_{i=1}^{N} \varepsilon 
%		\Big( {i\over \varepsilon} \langle \Phi_i(Z)| \Delta 
%		\Phi_i(Z) \rangle - \langle \Phi_i(Z)| H(t_i) | \Phi_i(Z) 
%		\rangle \Big)\Big \} \nonumber  \\
	&=& \int [d\mu(Z(t))] \exp \Big\{ i {\cal S} [Z(t)] \Big\},
\eea
where
\beq
	{\cal S} [Z(t)] = \int_{t_0}^{t_f} dt \Big\{ \langle \Phi(Z(t))| 
		i{d \over dt} | \Phi(Z(t)) \rangle - \langle \Phi(Z(t))| H(t) 
		| \Phi(Z(t)) \rangle \Big\}
\eeq
is an effective action in the generalized 
coherent state representation.
This path integral is defined over the coset space {\bf G/H}. The 
effective action contains two terms. The second term is the matrix 
element of Hamiltonian operator in the 
coherent states, which determines 
the static properties of the classical Hamiltonian. The first term 
is pure geometric, and it is indeed a Berry phase \cite{Sim83,Wil84}
that describes quantum fluctuations, and also determines the 
time-evolution of the system,
\beq
	\omega[{\bf G/H}] = \int_{\Gamma \in \bf G/H} \langle \Phi(Z)| 
		d |\Phi(Z)\rangle = \int_{\Gamma \in \bf G/H} {\cal A} 
		\cdot d\hat{\Omega},
\eeq
where ${\cal A}$ is a gauge vector 
potential defined over the coset space 
{\bf G/H}, and $\hat{\Omega}$ is a unit vector in {\bf G/H}. 
One can then define the gauge connection,
\beq
	F \equiv  \langle d\Phi(Z)|d \Phi(Z) \rangle) = 
		\sum_{\alpha \alpha'} \omega_{\alpha \alpha'} 
		dZ_\alpha \wedge dZ_{\alpha'}
\eeq
and $\omega_{\alpha \alpha'}$ is the Berry curvatures:
\beq
	\omega_{\alpha \alpha'} = \Big\langle {\partial \Phi(Z) 
		\over \partial Z_\alpha}\Big| {\partial \Phi(Z) \over
		\partial Z_{\alpha'}} \Big\rangle - \Big \langle 
		{\partial \Phi(Z) \over \partial Z_{\alpha'}}\Big| {\partial 
		\Phi(Z) \over \partial Z_\alpha} \Big\rangle .
\eeq
When the rank of {\bf G} is larger than one, the associated gauge 
potential ${\cal A}$ is non-abelian with gauge group $\leq$ {\bf H}.
From the above generalized coherent state path integral, one can
study the so-called geometric quantization \cite{Woo92,Fed89} and 
classical gauge equations of motion in quantum mechanics 
\cite{Won70,Bal91}. 

This path integral formulation  and the 
associated gauge potentials have potential applications in 
condensed matter physics and particle physics. This is because
classical semisimple Lie groups can be generated by bilinear 
operators of  bosonic and fermionic creation and annihilation 
operators. The bilinear operators describe the basic collective
excitations in strongly correlated or strongly coupled systems.  
Therefore, the above formalism can be applied 
directly to various realistic physical problems. 

Specifically, the SU(n) group can be generated by the 
particle-hole pairs: $\{a^\dagger_i a_j$; $1 \leq i,j \leq n \}$, 
and the corresponding generalized coherent state is given by
\beq	\label{csph}
	|\{Z_{ij}\}\rangle = {\cal N}(Z) \exp \Big\{ \sum_{ij} Z_{ij}
		a^\dagger_i a_j \Big\} | m \rangle  .
\eeq
where $a^\dagger_i, a_i$ can be either bosonic or fermionic
creation and annihilation operators, and $| m \rangle$ contains
$m$ particles in the lowest states, $m<n$. For bosonic system,
the coherent states are defined on the coset pace $\{ Z_{ij} \}
\in$ SU(n)/SU(n-1)$\times$U(1). For fermionic space, the coset 
space $\{ Z_{ij} \}$ is SU(n)/SU(n-m)$\times$U(m).
The spin coherent state of the Heisenberg model discussed in the
last section is a special case where the spin operators take the
form:
\beq
\vec{S}_i = {1\over 2} \sum_{\alpha \beta} a^\dagger_{i\alpha} 
		\vec{\sigma}_{\alpha \beta} a_{i\beta},
\eeq
$\sigma$ is the Pauli matrix, and $\alpha, \beta$ denote
the spin index of electrons. Let $Z_{ii}=\tan{\theta\over 2} 
e^{-i\varphi}$, the spin coherent state (\ref{scshm}) can
be reduced to the form of (\ref{csph}) with $Z_{ij}=0$ for 
$i \neq j$ and ${\cal N}(Z)=1/(1+|Z_{ii}|^2)^s$. 
Correspondingly, the geometrical space SU(n)/SU(n-m)$\times$U(m) 
is reduced to $\{ Z_{ii} \} \in \prod_i\otimes$SU$_i$(2)/U$_i$(1).

The Sp(2n+1) group can be realized by bosonic particle-particle
and particle-hole pairs: $\{a^\dagger_i, a_i, a^\dagger_i a^\dagger_j, 
a_ia_j, a^\dagger_i a_j-{1\over 2}\delta_{ij} \}$. The generalized 
coherent state of Sp(2n+1) is given by 
\beq
	|\{z_i, Z_{ij}\}\rangle = {\cal N}(Z) \exp \Big\{\sum_i
		(z_i a^\dagger_i - z^*_i a_i)\Big\} \exp\Big\{ \sum_{ij} 
		{1\over 2}Z_{ij} a^\dagger_i a^\dagger_j \Big\} | 0 \rangle .
\eeq
The corresponding coset space $\{ z_i, Z_{ij} \}$ is SP(2n+1)/U(n).
The squeezed coherent states discussed in Sec. V are only 
special cases of the above coherent state. 

The SO(2n) group can be generated by fermionic particle-particle
and particle-hole pairs $\{c^\dagger_i c^\dagger_j, c_ic_j, 
c^\dagger_i c_j-{1\over 2}\delta_{ij} \}$. Similarly, one can write 
down the most general coherent state for SO(2n):
\beq
	|\{Z_{ij}\}\rangle = {\cal N}(Z) \exp \Big\{ \sum_{ij} Z_{ij}
		c^\dagger_i c^\dagger_j \Big\} | 0 \rangle  ,
\eeq
and its geometrical space is the coset space $\{ Z_{ij} \} \in$ 
SO(2n)/U(n). A typical example of the above coherent states is the 
BCS superconducting state in which only special fermionic pairs,
i.e., Cooper pairs $c^\dagger_{k\uparrow} c^\dagger_{-k\downarrow}$,
are considered \cite{BCS57}:
\beq
|{\rm BCS} \rangle = {1\over \sqrt{1+|h_k|^2}} \exp \Big\{ \sum_k h_k
		c^\dagger_{k\uparrow} c^\dagger_{-k\downarrow} 
		\Big\} | 0 \rangle .
\eeq
Since $\{c^\dagger_{k\uparrow} c^\dagger_{-k\downarrow}, 
c_{-k\downarrow}c_{k\uparrow}, c^\dagger_{k\uparrow}c_{k\uparrow}
+c^\dagger_{k\downarrow}c_{k\downarrow}-1\}$ span a su(2) algebra,
the geometrical space of the above BCS states is indeed the same
as the spin coherent state in Heisenberg model, i.e., $\{ h_k \} \in
\prod_k\otimes$SU$_k$(2)/U$_k$(1). Therefore, the BCS state carries
a U(1) gauge degree of freedom. But physically, the superconductivity
is very different from the ferro and antiferro-magnetism.
This is because the Heisenberg model has a global spin rotational
symmetry, while the BCS Hamiltonian only has a global U(1)
symmetry. In the Heisenberg model, the spontaneously breaking of 
spin rotational symmetry leads to the spin-wave excitations 
which can be described by the Non-Lonear Sigma Model derived 
from the spin coherent state path integral, as we have 
discussed in the previous section. In the BCS theory,
the spontaneously breaking of the U(1) symmetry for the pairing 
coherence gives pair excitations which can be described by
Ginzberg-Landau theory. The Ginzberg-Landau theory should also 
be derivable from the path integral of BCS coherent states.

The above general fermionic pairing coherent states can also be 
applies to systems other than the conventional BCS superconductivity. 
For example, if I take the triplet pairs:
\beq
	\vec{T^+}(k) = {1\over 2} \sum_{\alpha \beta}c^\dagger_{k\alpha}
		(i\vec{\sigma}\sigma_2 )_{\alpha \beta} c^\dagger_{-k\beta} 
		~,~~
	 \vec{T}(k) = {1\over 2} \sum_{\alpha \beta}c_{-k\alpha}
		(-i\sigma_2 \vec{\sigma})_{\alpha \beta} c_{k\beta} 
\eeq
together with the charge and spin operators:
\beq
	Q(k)={1\over 2}\sum_\alpha(c^\dagger_{k\alpha}c_{k\alpha}
		+c^\dagger_{-k\alpha}c_{-p\alpha}) -1 ~,~~
	\vec{S}(p)={1\over 2} \sum_{\alpha \beta}(c^\dagger_{k\alpha}
		\vec{\sigma}_{\alpha \beta} c_{k\beta}+c^\dagger_{-k\alpha}
		\vec{\sigma}_{\alpha \beta} c_{-k\beta}),
\eeq
which generates a SO(5) group, I can construct a generalized 
coherent state for the triplet pairing for superfluid $^3$He:
\beq
	|{\rm SF} \rangle = {\cal N}(Z) \exp \sum\{ \vec{Z}_k \cdot 
		\vec{T}^\dagger (k)\Big\} |0\rangle . 
\eeq
Its coset space is $\{ \vec{Z}_k \} 
\in \prod_k\otimes$SO$_k$(5)/U$_k$(2). 
This SO(5) coherent state carries a non-abelian 
SU(2) gauge. One can use this SO(5) generalized coherent state to 
study non-abelian gauge fields and low energy effective theory 
for superfluid $^3$He atoms \cite{Leg75}. 
Recently, I also constructed 
a generalized pairing state to include the singlet and triplet 
pairs, 
\begin{eqnarray}
| {\rm ZW} \rangle &=& {\cal N}(Z){\prod}'_{\bf k}\exp \Big\{Z_1({\bf k})
		c^\dagger_{{\bf k} \uparrow}c^\dagger_{-{\bf k}\downarrow} 
		+ Z_2({\bf k})c^\dagger_{{\bf k}+{\bf Q} \uparrow}
		c^\dagger_{-{\bf k}+{\bf Q} \downarrow}  \nonumber \\
		& & ~~~~~~~~~ + Z_3({\bf k})c^\dagger_{{\bf k} \uparrow}
		c^\dagger_{-{\bf k}+{\bf Q}\downarrow} + Z_4({\bf k})
		c^\dagger_{{\bf k} \downarrow}c^\dagger_{-{\bf k}+{\bf Q}
		\uparrow} \nonumber \\
		& & ~~~~~~~~~ + Z_5({\bf k})c^\dagger_{{\bf k} \uparrow}
		c^\dagger_{-{\bf k}+{\bf Q}\uparrow}+ Z_6({\bf k})
		c^\dagger_{{\bf k} \downarrow}c^\dagger_{-{\bf k}+{\bf Q}
		\downarrow} \Big\} | 0 \rangle ,	\label{gcs}
\end{eqnarray}
for the study of high $T_c$ superconductivity and the close proximity 
between the Mott insulating antiferromagnetic order and $d$-wave 
superconducting order in cuprates\cite{Zha99}. 
Here the coset space is $\prod_k\otimes$SO$_k$(8)/U$_k$(4). 
Under the constraint of non-double occupied sites, possible gauge group 
contains in the above coherent pairing states may be SU(2)$\times$U(1) 
or a larger one up to U(4). This may open a new window for the study  
of the dynamical mechanism of high $T_c$ superconductivity.

If one takes the continuum limit in the coordinate space and lets 
$t_0 \rightarrow -\infty$, $t_f \rightarrow \infty$, then the 
path integral based on the generalized coherent states can be 
expressed as
\beq
G = \int [d\mu(\hat{\Omega}(x))] \exp \Big\{ i \int d^{d+1}x 
		\Big\{{\cal A}(x)\cdot \dot{\hat{\Omega}}(x) - 
		{\cal H}[\hat{\Omega}(x)]\Big\} ,
\eeq
where $x$ is a coordinate in the $d+1$ dimensional Minkowshi space.
If the Hamiltonian has a symmetry {\bf S} $\subset$ {\bf G}, and 
the static classical ground states (which can be obtained by
minimizing ${\cal H}$ with respect to the coherent state parameters)
spontaneously breaks this symmetry, then one can use the saddle-point
expansion to derive a general Non-Linear Sigma Model defined 
on {\bf G/H},
\beq
	G = \int [d\mu(\hat{\Omega}(x))] \exp \Big\{ i \int d^{d+1}x 
		\Big\{{1\over 2g} \partial_\mu \hat{\Omega(x)} \cdot
		\partial^\mu \hat{\Omega}(x) + \cdots + \Theta_{\rm Top} 
		\Big\} ,
\eeq
to describe the low energy physics in the long-wave length limit,
where $\{\cdots\}$ denotes the higher order derivatives in the 
Non-Linear Sigma Model, and $\Theta_{\rm Top}$ is a topological 
phase, corresponding to a Wess-Zumino-Witten topological term 
\cite{Wit84,Wie88,Sto89} that is induced by the gauge degrees 
of freedom contained in the generalized coherent states and/or
a Chern-Simons term of topological gauge fields over the
coset space {\bf G/H} \cite{Wit89}. There may exist many potential 
applications of such a Non-linear Sigma Model in real physical 
problems, such as quantum Hall effect \cite{Sto89,Zir99}, the high 
$T_c$ superconductivity \cite{Zha97,Dem99}, the disorder systems 
\cite{Efe96} in condensed matter physics. It is also possible to 
apply such theory to quantum chromodynamics in particle physics when 
quantum chromodynamics is formulated in lattices\cite{Wil75,Sha97}),
and quantum gravity \cite{Wit91}, etc.  

\section{Summation}
In summation, as I have emphasized throughout the article, 
coherent states possess three unique properties that are fundamental 
to field theory. The property of coherent behavior uniquely 
describes the processes involving infinite number of virtual particles. 
The coherent excitations obtained from coherent states also give 
the essential physical picture of long-range orders induced by strong 
correlations.  The property of overcompleteness provides a reformulation 
and generalization of the functional integral in field theory, in
which quantum fluctuations of composite operators are included
as new low energy dynamical field variables. One may thereby be 
able to determine the dynamical degrees of freedom in different 
energy scales and to derive the corresponding effective theory. 
The property of topologically nontrivial geometrical structure 
in generalized coherent states allows one to explore the origin 
of  gauge fields and associated gauge degrees of freedom.
In this article, I have not touched the recently development of 
coherent states in terms of superalgebras and quantum groups. 
These topics may also be very important in the modern development 
of field theory, such as in supersymmetry, superstring and 
conformal field theory. Nevertheless, in my personal opinions, 
understanding the origin as well as the nature of gauge degrees 
of freedom in physics is perhaps the most fundamental 
problem in field theory. 

%\begin{references}

%\end{references}

\end{document}